\documentclass{trbunofficial}
\usepackage{lineno}
\usepackage{amsmath}
\allowdisplaybreaks
\usepackage{breqn}
\usepackage{calc}  
\usepackage{enumitem}  
\usepackage{amsfonts}
\usepackage{mathrsfs}
\usepackage{algpseudocode}
\usepackage{xcolor}
\usepackage{algorithm}
\usepackage{natbib}
\usepackage{bigstrut}
\usepackage{rotating}
\usepackage{graphicx}
\usepackage{nomencl}
\usepackage{caption}
\usepackage{subcaption}
\usepackage{amsthm}
\usepackage{algorithm}
\usepackage{algpseudocode}
\newtheorem{theorem}{Theorem}
\newtheorem{lemma}{Lemma}
\newtheorem{corollary}{Corollary}

\usepackage{subcaption}
\usepackage{tikz}
\usepackage{graphicx, amsfonts,amsmath , boldline, epstopdf,amssymb }
\usepackage{tabu, multirow,multicol ,booktabs ,setspace ,algcompatible ,mathrsfs,dsfont, array, boldline, makecell, booktabs }
\usepackage{url}
\usepackage{epstopdf}
\usepackage{adjustbox, stackengine, color, colortbl, epsfig, cases, enumitem, mathtools, tikz, verbatim}
\usetikzlibrary{shapes,arrows.meta,decorations}
\definecolor{Abi}{rgb}{0.309803, 0.58039, 0.80392}
\definecolor{orange}{rgb}{1, 0.5019, 0}
\definecolor{Red}{rgb}{1, 0, 0}
\definecolor{orange}{rgb}{1, 0.50196, 0}
\definecolor{greenJ}{rgb}{0, 0.6590, 0.42}
\definecolor{Brown}{rgb}{0.588, 0.294, 0}
\definecolor{um}{rgb}{0.0824, 0.1294, 0.4196}
\definecolor{abikam}{rgb}{0.51, 0.93,  0.992}

\usepackage{tabularx}
\usepackage[hidelinks]{hyperref}

\AuthorHeaders{Rafi and Guo}
\title{\centering Multi-agent Optimization of Non-cooperative Multimodal Mobility Systems}

\author{%
  \textbf{Md Nafees Fuad Rafi}\\
  Graduate Research Assistant\\
  Department of Civil, Environmental and Construction Engineering \\
  University of Central Florida\\
  mdnafeesfuad.rafi@ucf.edu\\
  \hfill\break
  \textbf{Zhaomiao Guo, Ph.D.}\\
  Assistant Professor \\
  Maseeh Department of Civil, Architectural and Environmental Engineering \\
  The University of Texas at Austin\\
  zguo@utexas.edu\\
  \hfill\break%
}

\begin{document}
\maketitle
\section{Abstract}

While multimodal mobility systems have the potential to bring many benefits to travelers, drivers, the environment, and traffic congestion, such systems typically involve multiple non-cooperative decision-makers who may selfishly optimize their own objectives without considering the overall system benefits. This paper aims to investigate market-based interactions of travelers and ride-sourcing drivers in the context of multimodal mobility systems. We propose a unified mathematical modeling framework to capture the decentralized travelers and drivers' decision-making process and balance the network's demand and supply by equilibrium pricing. Such a model allows analyses of the impact of decentralized decision-making on multimodal mobility efficiencies. The proposed formulation can be further convexified to efficiently compute the equilibrium ride-sourcing prices. We conduct numerical experiments on different settings of transportation networks to gain policy insights. We find that travelers prefer ride-sourcing and multimodal transportation more than the driving option when they are more sensitive to prices. We also find that travelers may need to be subsidized to use multimodal transportation when there is fewer transit hubs in the network or, ride-sourcing drivers become too sensitive to the prices. However, we find that more transit hubs in the network increases the total empty VMT of ride-sourcing drivers by increasing the total relocation time. The proposed model can be used by policymakers and platform operators to design pricing and subsidy schemes that align individual decision-making with system-level efficiency and evaluate the trade-offs between accessibility and environmental impacts in multimodal transportation networks. 

\hfill\break%
\noindent\textit{Keywords}: Multimodal Transportation, Ride-sourcing, Spatial Pricing, Transit, Multi-agent Optimization
\newpage

\section{Introduction}
Multimodal Transportation (MT) systems integrate multiple transportation modes such as cars, buses, ride-sourcing, trains, bike-share, and walking to complete a single journey. MT has the potential to serve diverse travel demands, including the needs of travelers who cannot, should not, or prefer not to drive. For example, people may choose not to drive due to economic constraints, physical limitations, or safety concerns. MT systems can also reduce traffic congestion and environmental pollution by increasing vehicle occupancy and reducing car ownership. The occupancy rate of personal vehicles in the United States was only 1.5 travelers per vehicle in 2019 \cite{Center2023}, which is a major contributor to congestion and pollution in urban areas \cite{gu2021}. Moreover, \cite{Center2023} indicated that 24\% of U.S. households owned three or more vehicles in 2017, while cars and light-duty trucks together accounted for 58\% of U.S. transportation emissions and 16\% of total U.S. emissions. By offering economically viable travel alternatives and encouraging shared travel through transit or ridesharing, MT systems can help reduce car ownership, lower travel costs, and decrease emissions, contributing to a more equitable, accessible, and sustainable transportation future.

Ride-sourcing services provide an additional layer of flexibility and convenience within MT systems, particularly for bridging the first-mile/last-mile gap, which is critical to improving transit ridership \cite{BROWN2021100396}. These services offer efficient connections between travelers’ origins or destinations and transit stations, enhancing accessibility to the overall system. Previous studies have shown that ride-sourcing services are often complementary to transit. For instance, \cite{RAYLE2016168} suggested that ride-sourcing complements transit services despite competition for certain trips, while \cite{feigon2016shared} reported a positive correlation between shared mobility usage and increased transit use. A report by \cite{lyft2018} revealed that 21\% of Lyft users reported using public transit more frequently due to the availability of reliable first- and last-mile connections. Consequently, many transit agencies have started collaborating with ride-sourcing companies to improve user accessibility and provide better multimodal solutions \cite{ZHANG202195}. These observations highlight the importance of studying multimodal systems that include both ride-sourcing and transit as integral travel modes.

Several literatures have examined the value of MT and ride-sourcing services to improve mobility. For example, \cite{FRANK2021103111} presented a decision support tool for locating multimodal mobility hubs to improve intermodal accessibility to workplaces and points of interest. \cite{FENG2022103611} modeled order dispatching in MT systems, including transit and ride-sourcing, as a large-scale sequential decision-making problem and proposed both a reinforcement learning approach and an integer linear programming (ILP) model to optimally dispatch idle drivers and advise passengers on combined modes. Similarly, \cite{LIU2019648} applied a Bayesian optimization framework to determine the optimal supply-side parameters of Mobility-on-Demand (MoD) operations within MT systems, where travel demand depends on Level of Service (LOS). In addition, \cite{YAN2019683} used a joint Revealed-Preference/Stated-Preference model to show that ride-sourcing services providing convenient last-mile connections can enhance transit ridership. These studies primarily focus on improving MT system efficiency and coordination mechanisms between modes.

However, in practice, ride-sourcing providers are profit-driven and aim to maximize their own objectives rather than solely benefiting MT systems. This leads to competition between ride-sourcing and transit services, raising concerns about whether ride-sourcing ultimately complements or substitutes public transit \cite{ZHANG202195}. Moreover, coordinating different transportation services and road users in a multimodal environment is inherently challenging due to the presence of multiple self-interested agents. Each entity in the system acts autonomously, often prioritizing individual benefits over system-wide performance. To the best of our knowledge, prior research has not examined a multimodal system consisting of ride-sourcing, driving, and transit from a non-cooperative multi-agent perspective, nor has it addressed the challenge of designing spatial pricing mechanisms that balance travel flows across these modes.

To bridge these gaps, we propose a unified mathematical modeling framework for analyzing non-cooperative multimodal services. This framework captures the interactions among travelers, ride-sourcing drivers, and other travel modes, providing insights into how market competition can be influenced to promote sustainable transportation choices. Specifically, we develop a multi-agent optimization framework to model the multimodal mobility system and propose a novel pricing strategy for ride-sourcing services that balances locational supply and demand.

The main contribution of this research is as follows:
\begin{enumerate}
    \item We develop a multi-agent optimization framework to model the multimodal mobility systems and capture the non-cooperative behavior of travelers and drivers in the context of MT system. It captures the ride-sourcing behaviors as well as its interaction with driving and transit modes in a multimodal system.
    \item We propose an exact convex reformulation of the multi-agent problem, the solution of which not only satisfies the travelers and drivers objectives but also balances the drivers and travelers flow in a MT system.
    \item We identify the locational pricing of mobility services to balance locational mode-specific supply and demand.
\end{enumerate}

\section{Methodology} \label{sec:methodology}
In this section, we model the multimodal system by developing a multi-agent optimization framework. First, we develop mathematical models for capturing the choice of travelers and ride-sourcing drivers. Then, we introduce equilibrium conditions to balance the mode specific supply and demand in the MT system. The interactions of drivers and travelers' decision making will determine the equilibrium pricing of ride-sourcing services in both uni-modal and multi-modal systems.

\subsection{Problem Description}
We consider a transportation system with three available modes of transport for travelers: driving, ride-sourcing, and multimodal transportation (integrated mode of ride-sourcing and transit). The key stakeholders involved in this system are travelers, ride-sourcing drivers, Transportation Network Companies (TNCs), and transit providers. We assume that transit providers have fixed transit routes, scheduling and pricing strategies; TNCs aim to balance locational service supply and demand; and travelers and drivers aim to selfishly optimize their own objectives. In this multi-agent system, each stakeholder’s decision is made decentrally with consideration of their own objective. Though each stakeholder is independent in their decision making, the result of any individual stakeholder's decision making influence the others. They interact with each other until their decisions reach an equilibrium where no stakeholder can improve its benefit by unilaterally changing its actions. We aim to capture the complex interactions between travelers and ride-sourcing drivers and find the equilibrium locational ride-sourcing prices that balance this complex system.

Let's consider a city of $\mathcal{M}$ zones. These zones are connected by a transportation network, expressed as a directed graph $\mathcal{G} = (\mathcal{N}, \mathcal{A})$, where $\mathcal{N}$ denotes the set of transportation nodes or zones and $\mathcal{A}$ denotes the set of connection between nodes. $\mathcal{R}$ is the set of origins indexed by $r$, $\mathcal{S}$ is the set destinations indexed by $s$ and $\mathcal{H}$ is the set of all transit hubs in the network indexed by $h(r,s)$. All destinations $\mathcal{S}$ and  transit hub $\mathcal{H}$ can be considered as the drop off point of ride-sourcing vehicles. Therefore, we denote the ride-sourcing vehicles drop-off point as $\mathcal{S'} = \left \{\mathcal{S} \cup \{h (r,s), \forall r \in \mathcal{R},\forall s \in \mathcal{S}\}\right \}$ or,
$\mathcal{S'} = \left \{\mathcal{S} \cup \mathcal{H}\right \}$ indexed by $s'$. 

Hence, the travelers' origin-destination (OD) set is, $\mathcal{{RS}} = \{(r,s) \mid r \in \mathcal{R}, s \in \mathcal{S}, r\neq s\}$ and ride-sourcing drivers OD set is,  $\mathcal{\overline{RS}} = \left \{(r,s') \mid r \in \mathcal{R}, s' \in \mathcal{S} \cup \{h (r,s), \forall s \in \mathcal{S}\}\right \}$. 

We assign a specific transit hub $h(r,s) \in \mathcal{H}$ for any OD set $(r,s) \in \mathcal{RS}$ i.e. travelers using multimodal services will be dropped of at transit hub $h(r,s) \in \mathcal{H}$ if they start their journey form $r \in \mathcal{R}$ and their destination is $s \in \mathcal{S}$. Note that, multiple OD set can also be assigned with the same transit hub and it will not change our proposed model framework. 

Each trip is initiated by a traveler who requests a pickup from origin zone $r$. Upon receiving the request, the traveler is matched with an idle driver in the same zone if available. Otherwise, he/she will be matched with a driver from a different zone. After a driver is matched to the passenger, the driver picks up the passenger, and chooses the path with the shortest travel time from $r$ to $s$ (if passenger chooses ride-sourcing) or $r$ to $h(r,s)$ (if passenger chooses multi-modal). When a trip is complete, the driver can choose either to remain in the destination zone, or, relocate to a different zone to pick up passengers, or sign out from the system.

\subsection{Traveler's mode choice model}
Let's consider a set of mode choice options for the travelers to go from $r$ to $s$, denoted as $\mathcal{K}$. In this study, travelers choose between three modes from their origin $r$ to destination $s$: $\mathcal{K} = \{\text{1: Driving ($D$), 2: Ride-sourcing $R$, 3: Multimodal ($M$)}\}$. Here, multimodal transportation refers the integration of ride-sourcing and transit services where ride-sourcing provides the first-mile connection from origin $r$ to the transit hub $h(r,s)$. After arriving at the hub $h(r,s)$, travelers will take transit to reach the destination $s$. We adopt a multinomial logit model to capture travelers' mode choice behavior, with the general utility function specified in Equation \eqref{eq:traveler_utility}.

\begin{subequations}
\begin{align}
&&& {V_{rsk}^T} = {U_{rsk}^T} + \epsilon_{rsk}^T, \quad \ \forall (r,s) \in \mathcal{RS}, k \in \mathcal{K} \label{eq:traveler_utility}
\end{align}
\end{subequations}
The utility function $V_{rsk}^T$ of travelers denotes the utility of choosing mode $k$ to travel from origin $r$ to destination $s$. It consists of both deterministic component $U_{rsk}^T$ and random term $\epsilon_{rsk}^T$, which is assumed to follow a Gumbel distribution. $U_{rsk}^T$ depends on the attractiveness of mode $k$, travel time, and cost related to mode $k$, which can be written in Equation \eqref{utility:travelers}. 

\begin{subequations} \label{utility:travelers}
\begin{align} 
& {U_{rs1}^T} & &= \beta_{0,1}^T-\beta_{11}^T (t_{rs}+t_s^P)-\beta_{2}^T ({c_{rs}^C} + c_{s}^P), \quad \ \forall (r,s) \in \mathcal{RS} \label{eq:driving_utility} \\
& {U_{rs2}^T} & &= \beta_{0,2}^T-\beta_{12}^T{t_{rs}}-\beta_{2}^T {\eta_{rs}^R}, \quad \forall (r,s) \in \mathcal{RS} \label{eq:ride-sourcing_utility} \\
& {U_{rs3}^T} & &= \beta_{0,2}^T-\beta_{13}^T({t_{rh({r},{s})}}+{t_{h({r},{s})s}^B})-{\beta_{1}^{T''}} {w_{h({r},{s})s}^B}-\beta_{2}^T ({\eta_{rh({r},{s})}^R}+{\rho_{h({r},{s})s}^B}), \quad  \forall (r,s) \in \mathcal{RS}\label{eq:multimodal_utility}
\end{align}
\end{subequations}
\\
For driving utility, $t_{rs}$ is the driving time from $r$ to $s$ and $t_s^p$ is the parking time at destination $s$. $c_{rs}^c$ is the cost associated with traveling from $r$ to $s$ via driving and $c_s^p$ is the cost of parking at destination $s$. For ride-sourcing, $t_{rs}$ is the travel time from $r$ to $s$, which is identical to driving time. $\eta_{rs}^R$ is the amount of payment that a traveler has to pay to take the ride-sourcing service from $r$ to $s$. For multimodal transportation to go from $r$ to $s$, $t_{rh(r,s)}$ is the travel time from $r$ to transit hub $h$. $t_{h(r,s)s}$ is the transit travel time from transit hub $h(r,s)$ to destination $s$ and $w_{h(r,s)s}^T$ is the waiting time of a traveler at transit hub $h(r,s)$ for the transit vehicle to go to the destination $s$. $\eta_{rh(r,s)}^R$ is the amount of payment that a traveler has to pay to take the ride-sourcing service from $r$ to transit hub $h$. $\rho_{h(r,s)s}^B$ is the transit fare from $h$ to $s$. 

$\beta_{0,1}^T$, $\beta_{0,2}^T$, and $\beta_{0,3}^T$ are the coefficients representing the attractiveness of driving, ride-sourcing and multimodal mode for travelers. The rest of the $\beta$ coefficients represent the sensitivity of the corresponding parameters/variables.  Notice that since the travel time for different modes can be different (for example, people can better utilize their time as passengers than being a driver himself), the travel time coefficient $\beta_{1k}$ is mode $k$ specific. 

The probability of choosing any of the three available modes can be written as follows,
$$P_{rsk}^T = \frac{e^{{U_{rsk}^T}}}{{\sum_{k \in \mathcal{K}}e^{{U_{rsk}^T}}}}$$ 
Here, $P_{rsk}^T$ is the probability of travelers going from $r$ to $s$ using mode $k$. Hence, the flow of traffic from $r$ to $s$ using mode $k$ can be written as equation \eqref{2:rider mnl}, where, $d_{rs}^T$ is the total travel demand of travelers from $r$ to $s$ and $q_{rsk}^T$ is the flow of travelers from $r$ to $s$ using mode $k$.

\begin{equation} \label{2:rider mnl}
{q_{rsk}^T} = {d_{rs}^T} P_{rsk}^T 
\end{equation}
\\
As a starting point of developing an unified model for the MT system, we propose an equivalent convex optimization model for solving the traveler's mode choice model (Equation~\eqref{2:rider mnl}). The equivalent model for the rider's mode choice decision-making can be formulated as follows,\\
\begin{subequations} \label{3:rider model}
\begin{align}
& \underset{\boldsymbol{q}^T \geq 0} {\text{minimize}}
& &\sum_{(r,s) \in \mathcal{RS},k \in \mathcal{K}} q_{rsk}^T (\ln{q_{rsk}^T}-1-{U_{rsk}^T})
\label{obj:traveler} \\ 
& \text{subject to}
& &\sum_{k \in \mathcal{K}} q_{rsk}^T = d_{rs}^T \quad \forall (r,s) \in \mathcal{RS} 
\label{3a:rider flow balance_1}
\end{align}
\end{subequations}

In the objective function of the optimization model, the term $q_{rsk}^T \ln{q_{rsk}^T}$ corresponds to the entropy of the trip distribution. Though the rest of the objective function doesn't provide any intuitive physical interpretation, it is mathematically constructed in such a way that the first-order derivative of the Lagrangian function with respect to the decision variable $q_{rsk}^T$ results as the flow of travelers from the traveler's model choice model. We can determine the modal split between the three modes with \eqref{3:rider model}. Constraint \eqref{3a:rider flow balance_1} represents the balance of total travel demand through all available modes from $r$ to $s$. 

We can expand the traveler's optimization model \eqref{3:rider model} with the utility functions corresponding to each $k$ from Equations \eqref{utility:travelers} as follows, where all the notations have the same meaning as defined earlier.\\
\begin{subequations} \label{mod:travelers}
\begin{align}
& \underset{\boldsymbol{q}^T \geq 0} {\text{minimize}}
& & \sum_{(r,s) \in \mathcal{RS}} \{q_{rs1}^T [\ln{q_{rs1}^T}-1-\beta_{0,1}^T+\beta_{11}^T ({t_{rs}} + t_{s}^P)+\beta_{2}^T ({c_{rs}^C} + c_{s}^P)] \nonumber \\
&&& + q_{rs2}^T [\ln{q_{rs2}^T}-1-\beta_{0,2}^T+\beta_{12}^T ({t_{rs}})+\beta_{2}^T ({\eta_{rs}^R})] \nonumber \\
&&& + q_{rs3}^T [\ln{q_{rs3}^T}-1-\beta_{0,3}^T+\beta_{13}^T ({t_{rh({r},{s})}}+{t_{h({r},{s})s}^B})+\beta_{1}^{T''} ({w_{h({r},{s})s}^B})+\beta_{2}^T ({\eta_{rh({r},{s})}^R}+{\rho_{h({r},{s})s}^B})] \}\nonumber \\
\label{obj:traveler} \\ 
& \text{subject to}
& & \sum_{k \in \mathcal{K}} q_{rsk}^T = d_{rs}^T \quad \forall (r,s) \in \mathcal{RS}
\label{3a:rider flow balance}
\end{align}
\end{subequations}

\begin{lemma} \label{lemma 1}
The optimal solution (\textbf{$q^T$}) of problem \eqref{3:rider model} are the equilibrium solutions of the mode choice of travelers with multinomial logit model \eqref{2:rider mnl} given the ride-sourcing prices $\eta_{rs}^{R}$ and $\eta_{rh(r,s)}^{R}$.
\end{lemma}

\begin{proof}
See \cite{github_multimodal_optimization}
\end{proof}

Lemma~\ref{lemma 1} states that the optimization model \eqref{mod:travelers} and mode choice model \eqref{2:rider mnl} are mathematically equivalent, and both of them generate the same solution. However, since traveler's ride-sourcing payments $\eta_{rs}^R$ and $\eta_{rh({r},{s})}^R$ vary spatially depending on the locational supply of drivers and demand of travelers, we do not know these prices beforehand. Hence, the optimization model~\eqref{mod:travelers} cannot be directly solved in isolation, as the mode choice probabilities and resulting flows are endogenous to the system. 

\subsection{Driver's choice model}
Ride-sourcing drivers are sensitive to expected wage and relocation time to determine whether and where to provide ride-sourcing services. The expected wage is generally a function of the locational ride-sourcing prices that drivers receive from travelers. Generally, drivers may remain in his current zone or he may relocate to a different zone to pick up passengers depending on the utility. Drivers may also sign out from the system from their current zone at any time if the expected utility is lower than a threshold (e.g., minimum wages). The utility function for the drivers' choice can be written as \eqref{eq:driver_utility},
\begin{subequations} \label{eq:driver_utility}
\begin{align}
&&&{V_{nrs'}^D} = {U_{nrs'}^D} + \epsilon_{nrs'}^D  = \beta_{0,r}^{D}-\beta_{1}^Dt_{nr}+\beta_{3}^D \rho_{rs'}^R+\epsilon_{nrs'}^D, \quad \forall n \in \mathcal{N}, (r,s') \in \mathcal{\overline{RS}}\label{eq:relocation_utility}\\
&&&{V_{nH}^D} = {U_{nH}^D} + \epsilon_{nH}^D  = \beta_{0,H}^{D}+\epsilon_{nH}^D , \quad \forall n \in \mathcal{N} \label{eq:signout_utility}
\end{align}
\end{subequations}

Here, $V_{nrs'}^D$ is the utility of drivers currently located at $n$, relocate to $r$ to pick up travelers, and drop off at $s'$. When $n=r$, the drivers do not relocate, rather they pick up traveler from their current location. $V_{nrs'}^D$ is the utility of drivers to sign out from the system. $t_{nr}$ is the relocation time of drivers from location $n$ to travelers origin $r$. If the relocation time is large for a zone, driver's utility of moving from his current zone to that zone becomes smaller. $\rho_{rs'}^R$ is the ride-sourcing pricing they receive from travelers for ride-sourcing services from $r$ to $s$ or $h$. If drivers choose to remain in their current zone to pick up travelers, $t_{nr} = 0$. If drivers choose to sign out from the system, the deterministic portion of the utility function would become, $U_{nH}^D = \beta_{0,H}^D$. $\beta_{0,r}^D$ is the coefficient representing the attractiveness of location $r$ for drivers and $\beta_{0,H}^D$ represents the attractiveness of signing out from the system. The rest of the $\beta$s are the sensitivity of corresponding parameters/variables. $\epsilon_{nrs'}^D$ and $\epsilon_{nH}^D$ are random term, which is assumed to follow a Gumbel distribution. 

The probability of driver choices can be obtained as follows,
\begin{subequations}
\begin{align}
& & & P_{nrs'}^D = \frac{e^{U_{nrs'}}}{{\sum_{(i,j)  \in \mathcal{\overline{RS}
}}e^{U_{nij}}} + e^{U_{nH}}}, \quad \forall n \in \mathcal{N}, (r,s')  \in \mathcal{\overline{RS}} \\
& & & P_{nH}^D = \frac{e^{U_{nH}}}{{\sum_{(i,j)  \in \mathcal{\overline{RS}
}}e^{U_{nij}}} + e^{U_{nH}}}, \quad \forall n \in \mathcal{N}, (r,s')  \in \mathcal{\overline{RS}}
\label{eq:driver_probability}
\end{align}
\end{subequations}
Here, $P_{nrs'}^D$ is the probability of the driver currently located at $n$ to relocate to $r$ in order to pick up travelers and drop them off at $s'$.

Hence, the flow of driver relocate from location $n$ to $r$ to pick up travelers and drop off at $s'$ (denoted as $q_{nrs'}^D$) and the rate of drivers signing out from location $n$ (denoted as $q_{nH}^D$) can be calculated from \eqref{eq:driver_flow1} and \eqref{eq:driver_flow2}, respectively, where $Q_{n} $ is the total number of drivers available at location $n$.
\begin{subequations}
\begin{align}
& & &q_{nrs'}^D = P_{nrs'}^D \cdot Q_{n} \quad \forall n \in \mathcal{N}, (r,s')  \in \mathcal{\overline{RS}} \label{eq:driver_flow1}\\
& & &q_{nH}^D =P_{nH}^D \cdot Q_{n} \quad \forall n \in \mathcal{N}, (r,s')  \in \mathcal{\overline{RS}} \label{eq:driver_flow2}
\end{align}
\end{subequations}


With the objective of developing an unified model for MT system, we propose an equivalent convex optimization model of the driver's choice model as follows,
\begin{subequations} \label{mod:driver}
\begin{align} 
&\underset{\boldsymbol{q}^D, \boldsymbol{Q}_n \geq 0}{\text{minimize}}
& & \sum_{n \in \mathcal{N}} [\sum_{(r,s') \in \mathcal{\overline{RS}}} q_{nrs'}^D (\ln{q_{nrs'}^D}-1-\beta_{0,r}^{D}+\beta_{1}^Dt_{nr}-\beta_{3}^D \rho_{rs'}^R) \nonumber \\
& & &+ q_{nH}^D (\ln{q_{nH}^D}-1-\beta_{0,H}^{D})] - \sum_{n \in\mathcal{N}}\lambda_{n} (Q_n-\Delta Q_n^D) 
\label{mod:driver obj} \\ 
& \text{subject to}
& & \sum_{(r,s') \in \mathcal{\overline{RS}}} {q_{nrs'}^D} + q_{nH}^D = Q_n \quad \forall n \in \mathcal{N}\label{cons:driver_flow_balance_1}
\end{align}
\end{subequations}

The decision variables here are $q^D$ and $Q_n$. The model \eqref{mod:driver} is mathematically constructed in such a way that the first order derivative of the Lagrangian function with respect to the decision variables $q_{nrs'}^D$, $q_{nH}^D$ and $Q_n$ would result same as the result of the driver's logit choice model. $\Delta Q_{n}^D$ is the flow rate of new drivers signing in at $n$, which is an exogenous parameter.  $\lambda_n$ is the value/penalty of an additional driver at location $n$. More insights about $\lambda_n$ are discussed in the later sections. Constraint \eqref{cons:driver_flow_balance_1} represents the drivers flow balance, i.e. the number of drivers available at location $n$ should be equal to the sum of relocation flow of drivers from $n$ to $r$ and the rate of drivers signing off from $n$. 

\begin{lemma} \label{lemma 2}
The optimal solutions (\textbf{$q^D, Q_n$}) of problem \eqref{mod:driver} are the equilibrium solutions of the choice decision of drivers with multinomial logit model \eqref{eq:driver_flow1} and \eqref{eq:driver_flow2} given the driver's ride sourcing prices $\rho_{rs'}^{R}$.
\end{lemma}

\begin{proof}
See \cite{github_multimodal_optimization}
\end{proof}

Lemma~\ref{lemma 2} states that the optimization model \eqref{mod:driver} and, driver's choice model \eqref{eq:driver_flow1} and \eqref{eq:driver_flow2}, are mathematically equivalent, and they generate the same solution. Similar to the traveler's model, we do not know the ride-sourcing prices $\rho_{rs'}^R$ beforehand, as it depends on the locational supply and demand. Hence, ride-sourcing driver's optimization model \eqref{mod:driver} cannot be used in isolation to determine the driver's flow across regions.

\subsection{Equilibrium conditions}
For multimodal system to be in a steady state, certain market clearing flow balance conditions need to be satisfied. We develop the following equilibrium conditions to balance the MT system.

\begin{subequations} \label{eq:equilibrium}
\begin{align} 
& & & ({{\rho_{rs}^R}}) & & \sum_{n \in \mathcal{N}} q_{nrs}^{D} = q_{rs2}^T  \quad \forall (r,s) \in \mathcal{RS} \label{ride source flow to destination driver and rider balance_1}\\
& & & ({{\rho_{rh(r,s)}^R}}) & & \sum_{n \in \mathcal{N}} q_{nrh(r,s)}^D = q_{rs3}^T  \quad \forall (r,s) \in \mathcal{RS} \label{ride source flow to hub driver and rider balance_1}\\
& & & (\lambda_{n}) & & Q_n = \sum_{r \in \mathcal{R}} q_{rn2}^T + \Delta Q^D_n \quad \forall n \in \mathcal
S\label{equilibrium constraint 3_1}\\
& & & (\lambda_{n}) & & Q_n = \sum_{(r,s) \in \mathcal{RS}(n)} q_{rs3}^T + \Delta Q^D_n \quad \forall n \in \mathcal
H\label{equilibrium constraint 4_1}\\
& & & (\lambda_{n}) & & Q_n = \Delta Q^D_n \quad \forall n \in \mathcal {N \setminus (S \cup H)
}\label{equilibrium constraint 5_1}
\end{align}
\end{subequations}

The flow of drivers relocating from $n$ to $r$ to pick up travelers and drop off at $s$ and $h(r,s)$ should be equal to the flow of travelers going from $r$ to $s$ taking ride-sourcing and multimodal modes, respectively. These conditions are satisfied through constraints \eqref{ride source flow to destination driver and rider balance_1} and \eqref{ride source flow to hub driver and rider balance_1}, respectively. The ride-sourcing prices $\rho_{rs}^R$ and $\rho_{rh(r,s)}^R$ can be interpreted as the dual variables of equilibrium constraints \eqref{ride source flow to destination driver and rider balance_1} and \eqref{ride source flow to hub driver and rider balance_1}, respectively. Here, $\rho_{rs}^R$ is the payment that drivers earn from giving ride-sourcing services to travelers going from $r$ to $s$ taking ride-sourcing directly. $\rho_{rh(r,s)}^R$ is the payment that a driver earns from giving ride-sourcing services to travelers going from $r$ to $s$ taking multimodal option. Constraint \eqref{equilibrium constraint 3_1} represents that the drivers available at destination $s$ is equal to the sum of the flow of travelers arriving at $s$ using ride-sourcing services from all origin $r$ and the number of new drivers newly logged into the system at $s$. The dual variable of this constraint $ \lambda_s,  \forall s \in \mathcal{S}$. $\lambda_s$ can be interpreted as the penalty or value of adding one additional driver at location $s$. Similarly, constraint \eqref{equilibrium constraint 4_1} represents that the drivers available at transit hub $h(r,s)$ is equal to the sum of the flow of travelers going from $r$ to $s$ via multimodal transportation and the number of new drivers newly logged into the system at transit hub $h(r,s)$. The dual variable of this constraint is $\lambda_{h(r,s)},   \forall h \in \mathcal{H}$. $\lambda_{h(r,s)}$ can be interpreted as the penalty or value of an additional driver at transit hub $h(r,s)$.  Constraint \eqref{equilibrium constraint 5_1} represents that drivers available at location $n$ which is neither a traveler's destination nor a transit hub are equal to the number of new drivers newly logged into the system at that location $n$ only. As no travelers go to a location which is not a destination or hub, no driver is going to end up at that location. The corresponding dual variable $\lambda_n \forall n \in \mathcal {N \setminus (S \cup H)}$ of this constraint can be interpreted as the penalty or value of adding one extra driver at any location $n$ that is neither a traveler's destination nor a transit hub.

\subsection{Solution Approach} \label{sec:Solution Approach}
In the multi-agent system, the decision process of one agent depends on the decision of other agents \cite{GUO2021}. Hence, the ride-sourcing prices in multimodal system is also determined by the collective actions of all the available agents, which for our study are the travelers and drivers. These interactions are described in the equilibrium conditions \eqref{eq:equilibrium}. Since both travelers and drivers model are convex optimization problem, one can reformulate both problems as sufficient and necessary complementarity problems (CPs) and solve both CPs together (e.g., \cite{GUO2017}). But solving the CPs directly may raise the problem of non-convexity and high dimensionality. To solve this, we develop a single convex optimization model as follows, which will yield the same pricing decision to balance locational ride-sourcing supply and demand. The computational contribution of this research is to propose an exact convex reformulation for the traveler's and driver's problems, which is formally stated in Theorem~\ref{theorem 1}.

\begin{theorem} \label{theorem 1}
As both traveler's model and driver's model are convex, the equilibrium states of agents' interactions in a perfectly competitive market i.e. \eqref{mod:travelers}, \eqref{mod:driver} and equilibrium conditions \eqref{eq:equilibrium} are equivalent to solving a single level convex optimization problem, formulated as \eqref{mod:reformulation}.
\begin{subequations} \label{mod:reformulation}
\begin{align}
& \underset{\boldsymbol{q}^T, \boldsymbol{q}^D, \boldsymbol{Q}_n \geq 0} {\text{minimize}}
& & \frac{1}{\beta_2^T}\sum_{(r,s) \in \mathcal{RS}} \{q_{rs1}^T [\ln{q_{rs1}^T}-1-\beta_{0,1}^T+\beta_{11}^T ({t_{rs}} + t_{s}^P)+\beta_{2}^T ({c_{rs}^C} + c_{s}^P)] \nonumber \\
&&& + q_{rs2}^T [\ln{q_{rs2}^T}-1-\beta_{0,2}^T+\beta_{12}^T ({t_{rs}})] \nonumber \\
&&& + q_{rs3}^T [\ln{q_{rs3}^T}-1-\beta_{0,3}^T+\beta_{13}^T ({t_{rh({r},{s})}}+{t_{h({r},{s})s}^B})+\beta_{1}^{T^{''}} ({w_{h({r},{s})s}^B})+\beta_{2}^T ({\rho_{h({r},{s})s}^B})] \} \nonumber\\
&&&+  \frac{1}{\beta_3^D}\sum_{n \in \mathcal{N}} [\sum_{(r,s') \in \mathcal{\overline{RS}}} q_{nrs'}^D (\ln{q_{nrs'}^D}-1-\beta_{0,r}^{D}+\beta_{1}^Dt_{nr})+ q_{nH}^D (\ln{q_{nH}^D}-1-\beta_{0,H}^{D})]
\label{obj:combined} \\ 
& \text{subject to}
& & \sum_{k \in \mathcal{K}} q_{rsk}^T = d_{rs}^T \quad \forall (r,s) \in \mathcal{RS} \label{cons:traveler_flow_balance}\\
& & & \sum_{(r,s') \in \mathcal{\overline{RS}}} {q_{nrs'}^D} + q_{nH}^D = Q_n \quad \forall n \in \mathcal{N}\label{cons:driver_flow_balance}\\
& ({{\rho_{rs}^R}}) & & \sum_{n \in \mathcal{N}} q_{nrs}^{D} = q_{rs2}^T  \quad \forall (r,s) \in \mathcal RS \label{ride source flow to destination driver and rider balance}\\
& ({{\rho_{rh(r,s)}^R}}) & & \sum_{n \in \mathcal{N}} q_{nrh(r,s)}^D = q_{rs3}^T  \quad \forall (r,s) \in \mathcal RS \label{ride source flow to hub driver and rider balance}\\
& (\lambda_{n}) & & Q_n = \sum_{r \in \mathcal{R}} q_{rn2}^T + \Delta Q^D_n \quad \forall n \in \mathcal{S} \label{equilibrium constraint 3}\\
& (\lambda_{n}) & & Q_n = \sum_{(r,s) \in \mathcal{RS}(n)} q_{rs3}^T + \Delta Q^D_n \quad \forall n \in \mathcal
H\label{equilibrium constraint 4}\\
& (\lambda_{n}) & & Q_n = \Delta Q^D_n \quad \forall n \in \mathcal {N \setminus (S \cup H)
}\label{equilibrium constraint 5}
\end{align}
\end{subequations}
\end{theorem}

\begin{proof}
See \cite{github_multimodal_optimization}
\end{proof}

Using Theorem~\ref{theorem 1}, we can solve both travelers and drivers optimization model simultaneously and determine the equilibrium ride-sourcing prices. The main intuition behind Theorem~\ref{theorem 1} is the reverse procedure of Lagrangian relaxation, where we move the $q_{rs2}^{T} (\rho_{rs}^R + \lambda_s)$, $q_{rs3}^{T} (\rho_{rh(r,s)}^R + \lambda_{h(r,s)})$ terms in the traveler's model and $q_{nrs'}^D \rho_{rs'}^R, \lambda_{n} (Q_n-\Delta Q_n^D)$ terms in the driver's model from the objective functions to the constraints. Therefore, the objective function here is a combination of scaled objective function from the traveler's and driver's models except the $q_{rs2}^{T} (\rho_{rs}^R + \lambda_s)$,  $q_{rs3}^{T} (\rho_{rh(r,s)}^R + \lambda_{h(r,s)})$ terms from traveler's model and $q_{nrs'}^D \rho_{rs'}^R$, $\lambda_{n} (Q_n-\Delta Q_n^D)$ terms from the driver's model. The purpose of this scaling is to convert the objective function of both traveler's and driver's models into monetary units so that they are addable and we can recover the price portion of travel ($\rho_{rs}^R , \lambda_s$ or $\rho_{rh(r,s)}^R , \lambda_{h(r,s)})$ as the dual variable of the equilibrium constraints. By solving the single-level exact convex reformulation problem \eqref{mod:reformulation}, we will find the travelers flow, drivers flow, and total drivers availability in different zones, as well as the ride-souring pricing in the MT system.

\begin{corollary} \label{corollary 1}
The reformulation model \eqref{mod:reformulation} is separable to traveler's optimization model \eqref{mod:travelers} and driver's optimization model \eqref{mod:driver} if the following conditions hold.
\begin{subequations}
\begin{align}
& & &\eta_{rs}^R = \rho_{rs}^R + \lambda_s \quad \forall (r,s) \in \mathcal{RS} \label{tr_ride=dr_ride+val}\\
& & &\eta_{rh(r,s)}^R = \rho_{rh(r,s)}^R + \lambda_{h(r,s)} \quad \forall (r,s) \in \mathcal{RS} \label{tr_multi=dr_ride+val}
\end{align}
\end{subequations}
\end{corollary}

\begin{proof}
See \cite{github_multimodal_optimization}
\end{proof}

\begin{corollary} \label{corollary 2}
As traveler's optimization model \eqref{mod:travelers} and driver's optimization model \eqref{mod:driver} are strictly convex functions, the system equilibrium exists and is unique.
\end{corollary}

\begin{proof}
See \cite{github_multimodal_optimization}
\end{proof}
Corollary~\ref{corollary 1} implies that the traveler's price through ride-sourcing for any OD $(r,s)$ is the summation of the driver's revenue through ride-sourcing for OD $(r,s)$ and the value/penalty corresponding to an additional driver at the destination $s$. Also, the traveler's price through multimodal transportation for any OD ($r,s$) is the summation of the driver's price serving the multimodal mode through ride-sourcing service for OD ($r,h(r,s)$) and the value/penalty corresponding to the addition of one extra driver at the transit hub $h(r,s)$.

\section{Numerical Examples} \label{sec:numerical examples}
In this section, we test our models and solution approaches using a 5-node test network (Figure \ref{fig:Test Network}) and the Sioux Falls Network (Figure~\ref{fig:flow15}) to gain insights from our model. We implement our model using Python Optimization Modeling Objects: Pyomo v6.7.1 \cite{bynum2021pyomo}, \cite{hart2011pyomo}. For solving the numerical problems, we used the IPOPT 3.14.16 \cite{wachter2006} solver. All tests were run using a 2.80 GHz 11th Generation Intel Core i7 processor with 16 GB of RAM under a 64-bit Windows 11 operating system.

\subsection{5 Node Test Network}
\begin{figure}
  \centering
  \includegraphics[width=0.9\textwidth]{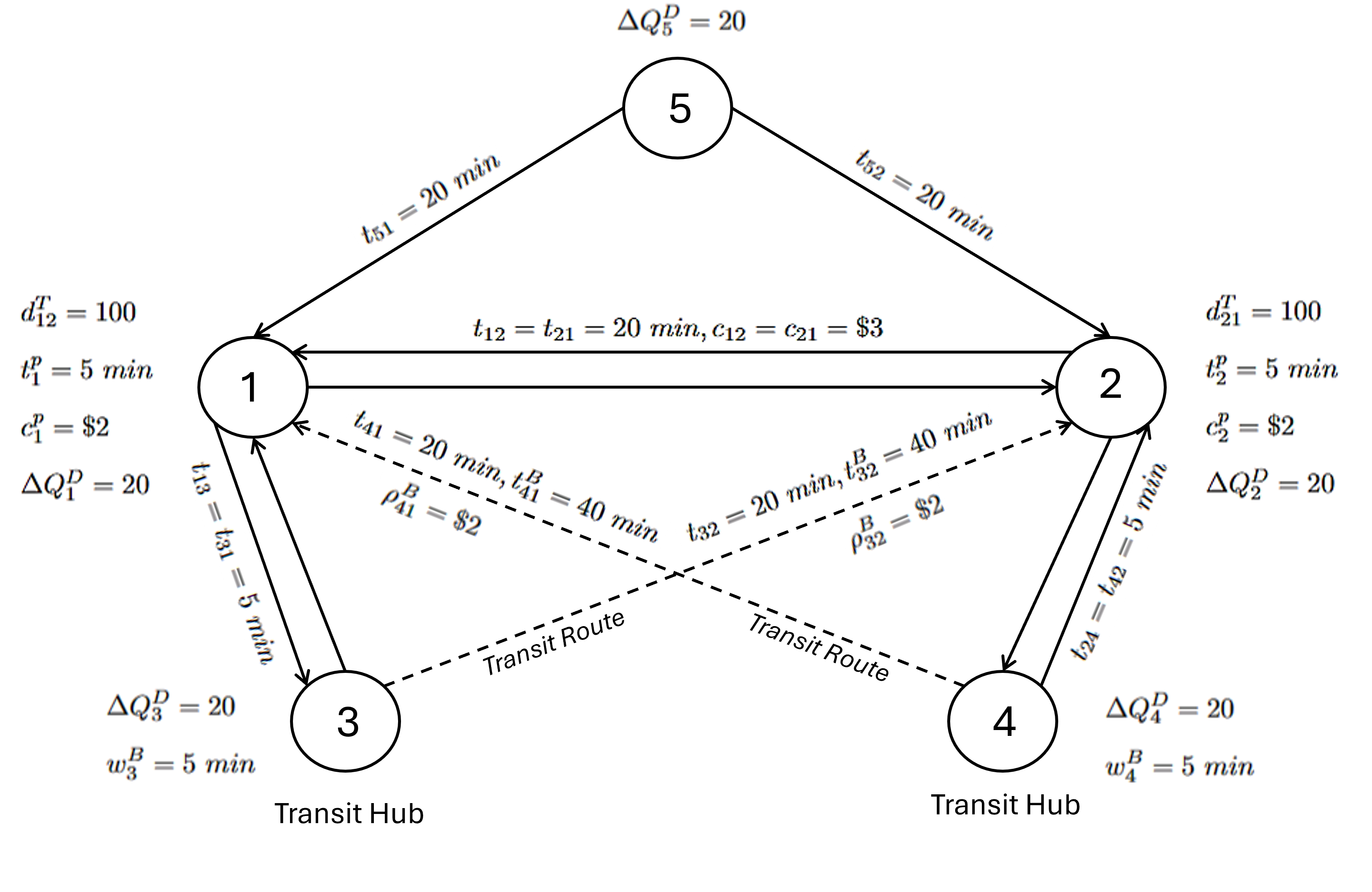}
  \caption{5 Node Test Network}
  \label{fig:Test Network}
\end{figure}

This network (Figure \ref{fig:Test Network}) consists of 5 nodes, $\mathcal{N}=[1,2,3,4,5]$. There are 10 available links connecting the nodes, i.e., $\mathcal{A}= [(1,2),(2,1),(1,3),(3,2),(2,4),(4,1),(3,1),(4,2),(5,1),(5,2)]$. It has two origin nodes $\mathcal{R}=[1,2]$ and two destination nodes $\mathcal{S}=[2,1]$ for travelers.Traveler's OD pairs, $\mathcal{RS}=[(1,2), (2,1)]$. Nodes 3 and 4 will act as the transit hub for travelers OD pair (1,2) and (2,1), respectively. Hence, we have transit hub set $\mathcal{H}=[3,4]$, where, $h(1,2)=3$ and $h(2,1)=4$. Note that any single transit hub may also be used by multiple OD pairs and it will not change our modeling structure. Node 5 in the network is neither a traveler's origin /destination nor a transit hub. Drivers can log into the system in any of the nodes $\mathcal{N}$. They will decide whether to stay in their current node, or relocate to any origin nodes of travelers, or sign out from the system based on the corresponding utility for each option. If drivers stay or relocate, they pick up travelers from the origin nodes and drop them at their destinations or transit hubs. Hence, drivers' OD pair from the origin node, $\mathcal{\overline{RS}} = [(1,2),(2,1),(1,3),(2,4)]$. The transit routes are considered along links (3,2) and (4,1).

Some initial model parameters are given in Figure \ref{fig:Test Network}. We consider an equal travel demand of 100 for both of the traveler's OD pair (1,2) and (2,1). We consider that 20 drivers logged into each node of the transportation network during the studied time horizon. The travel time for each of the links is considered as exogenous variable in our model. For links (3,2) and (4,1), transit travel time will be higher than driving time (40 minutes for transit and 20 minutes for driving). Notice that this 5-node test instance is intentionally kept as symmetric initially through which the model behavior can be understood better. Later, we will make the network asymmetric by simply changing the initial value of one parameter/coefficient to evaluate the model performance for different ranges of values of parameters/coefficients. The initial values of the sensitivity coefficients are $\beta_{0,1}^T$ = 4.0, $\beta_{0,2}^T$ = 2.0 , $\beta_{0,3}^T$ = 1.0, $\beta_{0,H}^D$ = 2.0, $\beta_{0,r}^D$ = 0.0, $\beta_{11}^T$ = 0.3, $\beta_{12}^T$ = 0.2, $\beta_{13}^T$ = 0.1, $\beta_{2}^T$ = 1.0, $\beta_{1}^{T''}$ = 0.2, $\beta_{2}^D$ = 0.2, $\beta_{1}^D$ = 0.3, $\beta_{3}^D$ = 1.0.

When we solve the convex reformulation model \eqref{mod:reformulation}, we get the traveler's flow ($q^T$), driver's flow ($q^D$), number of available drivers ($Q_n$), and travelers and drivers ride-sourcing prices corresponding to ride-sourcing and multimodal transportation. The methods mentioned previously solve this model efficiently in just 0.10 seconds. The traveler's flow found from this model was validated using traveler's convex optimization model \eqref{mod:travelers} and traveler's multinomial logit model \eqref{2:rider mnl} with the estimated ride-sourcing prices. The flow for each of these models was the same up to twelve decimal points. Similarly, driver's flow ($q^D$) and the number of available drivers ($Q_n$) were validated using the driver's convex optimization model \eqref{mod:driver} driver's multinomial logit choice model \eqref{eq:driver_flow1} and \eqref{eq:driver_flow2} given the drivers ride-sourcing prices $\rho_{rs'}$ estimated from the reformulation model \eqref{mod:reformulation}. Driver's flow and number of available drivers were the same up to twelve decimal points in all those models, evidencing that our reformulation is correct and the solution is unique and exact.  

\begin{figure}
  \centering
  \includegraphics[width=0.6\textwidth]{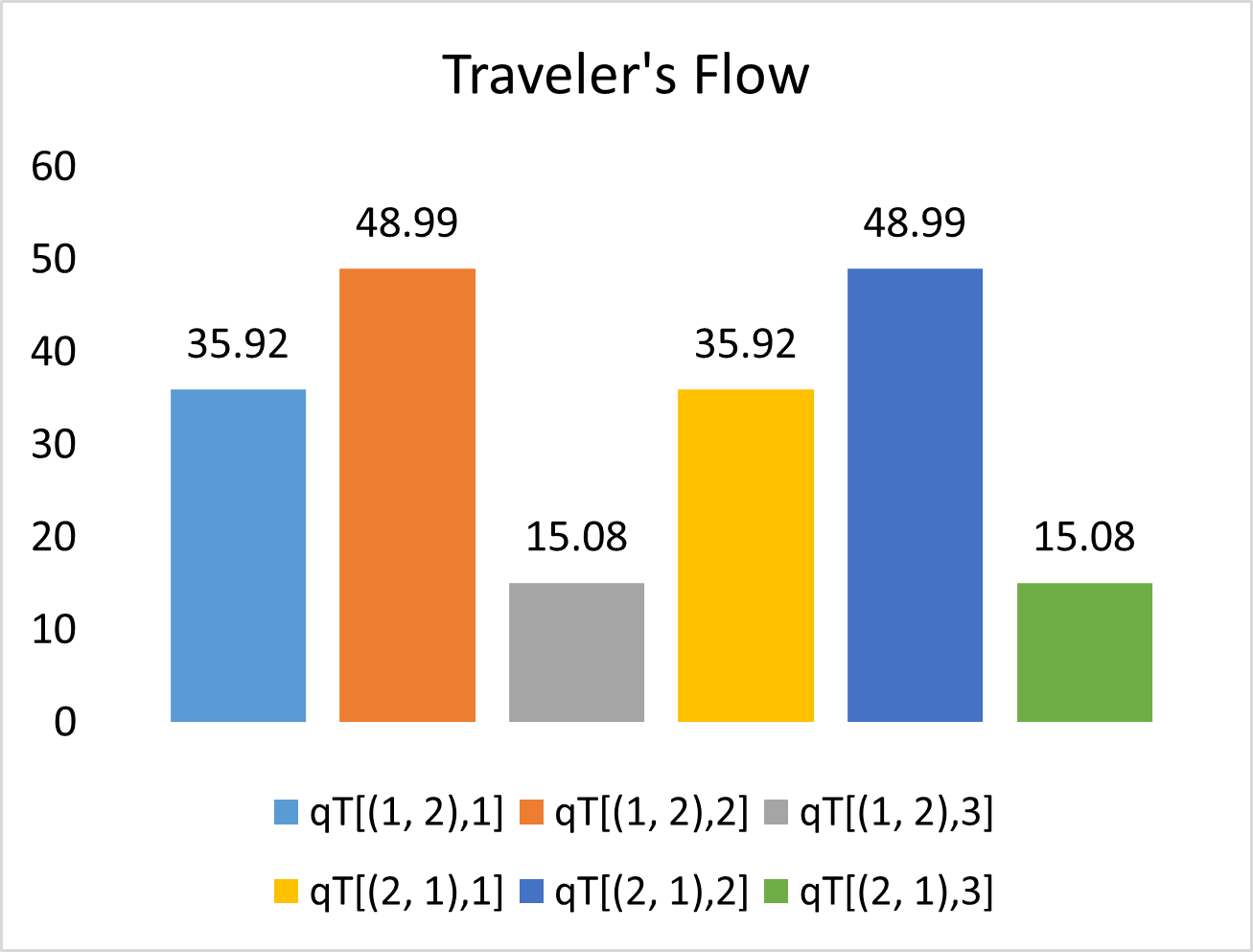}
  \caption{Traveler's Flow}
  \label{fig:Traveler's Flow}
\end{figure}

\begin{figure}
  \centering
  \includegraphics[width=1.0\textwidth]{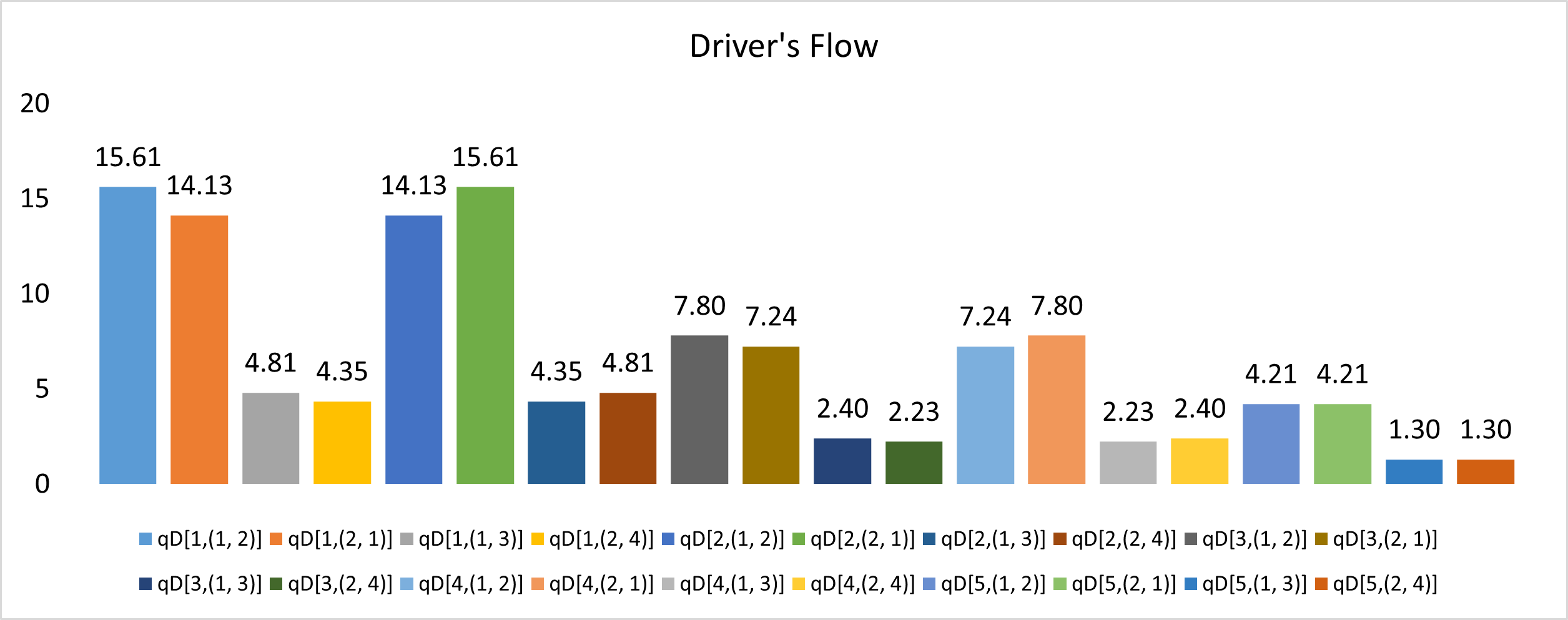}
  \caption{Driver's Flow}
  \label{fig:Driver's Flow}
\end{figure}

\begin{figure}[ht]
    \centering
    \begin{subfigure}[b]{0.49\textwidth}
        \centering
        \includegraphics[width=\textwidth]{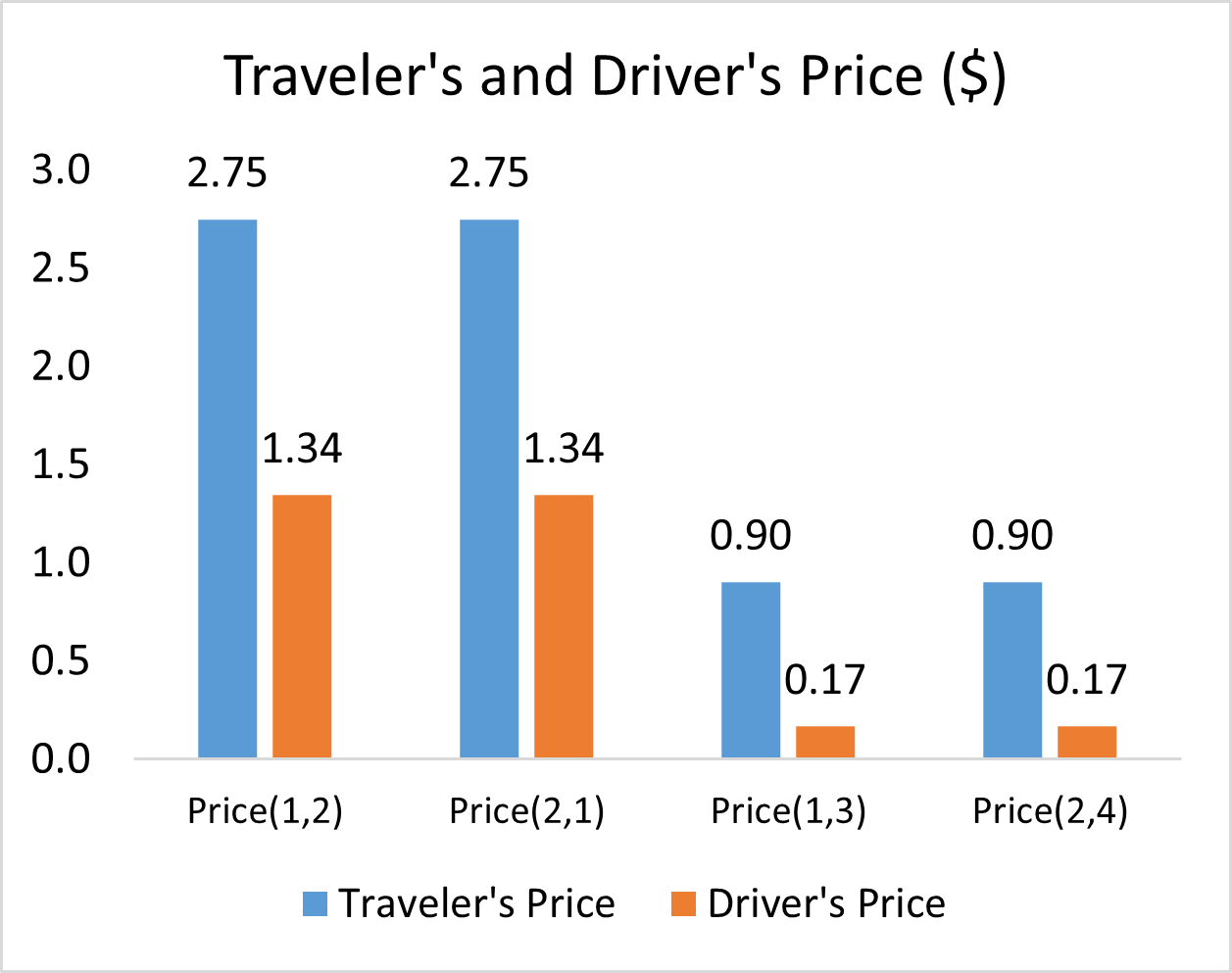}
        \caption{Traveler's and Driver's Price}
        \label{fig:Traveler's and Driver's Price}
    \end{subfigure}
    \hfill
    \begin{subfigure}[b]{0.49\textwidth}
        \centering
        \includegraphics[width=\textwidth]{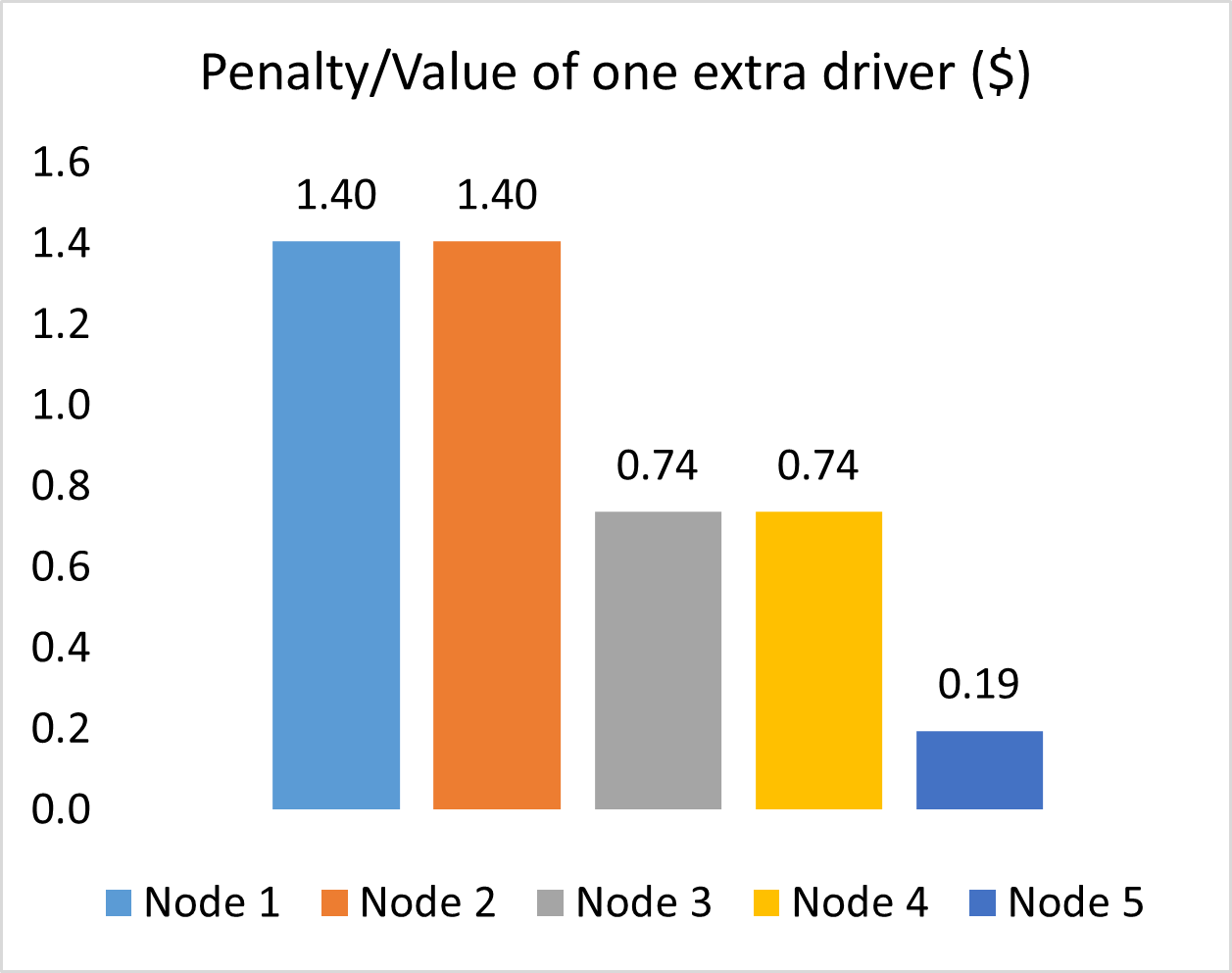}
        \caption{Value of one extra driver}
        \label{fig:Value of one extra driver}
    \end{subfigure}
    \caption{Traveler's and Driver's Price and Value of an additional driver}
    \label{fig:flow2}
\end{figure}

In Figure \ref{fig:Traveler's Flow}, the same traveler's flow ($q^T$) is found for OD (1,2) and (2,1) because the transportation network was symmetric. Figure \ref{fig:Traveler's Flow} shows that most travelers take the ride-sourcing option, followed by driving and multimodal transportation. Though the driving option has higher mode attractiveness, travelers' flow through driving is lower than ride-sourcing flow in our case. This can be explained by comparing the traveler's ride-sourcing prices in Figure \ref{fig:Traveler's and Driver's Price} with the driving cost. As travel time is equal for both driving and ride-sourcing, cost becomes the deciding factor for mode choice. For a specific OD, the total driving cost is higher than ride-sourcing cost at equilibrium. Though multimodal transportation has cheaper prices as shown in Figure \ref{fig:Traveler's and Driver's Price}, travelers prefer it less because of higher travel and waiting times at transit hubs. Additionally, people are more sensitive to waiting time and hence, the waiting time at the transit hub makes the total perceived travel time higher than those of the other two modes. Furthermore, few drivers are willing to serve the multimodal mode as evident from Figure \ref{fig:Driver's Flow}. As drivers' supply is less, the demand for multimodal service becomes smaller to balance the system and reach equilibrium.

Figure \ref{fig:Driver's Flow} shows that drivers at travelers' origin nodes tend to stay there to pick up travelers rather than relocate. For example, drivers' flows of $q^D[1,(1, 2)]$ and $q^D[2,(2, 1)]$ are higher than relocation flows. One reason is that drivers are sensitive to relocation time and ride-sourcing prices. The higher the relocation time and the lower the price at a location, the less likely drivers will relocate. Additionally, driver flow serving multimodal services is lower than that serving direct ride-sourcing. Usually, transit hubs are close to the traveler's origin, offering lower ride-sourcing prices (i.e., low wages) for drivers. Hence, fewer drivers serve multimodal service, discouraging travelers' decision to choose it, as observed in Figure \ref{fig:Traveler's Flow}. Travelers' ride-sourcing and multimodal flow and driver flow mutually influence one another until equilibrium satisfying conditions \eqref{eq:equilibrium} is reached.

\begin{figure}[ht]
    \centering
    \begin{subfigure}[b]{0.49\textwidth}
        \centering
        \includegraphics[width=\textwidth]{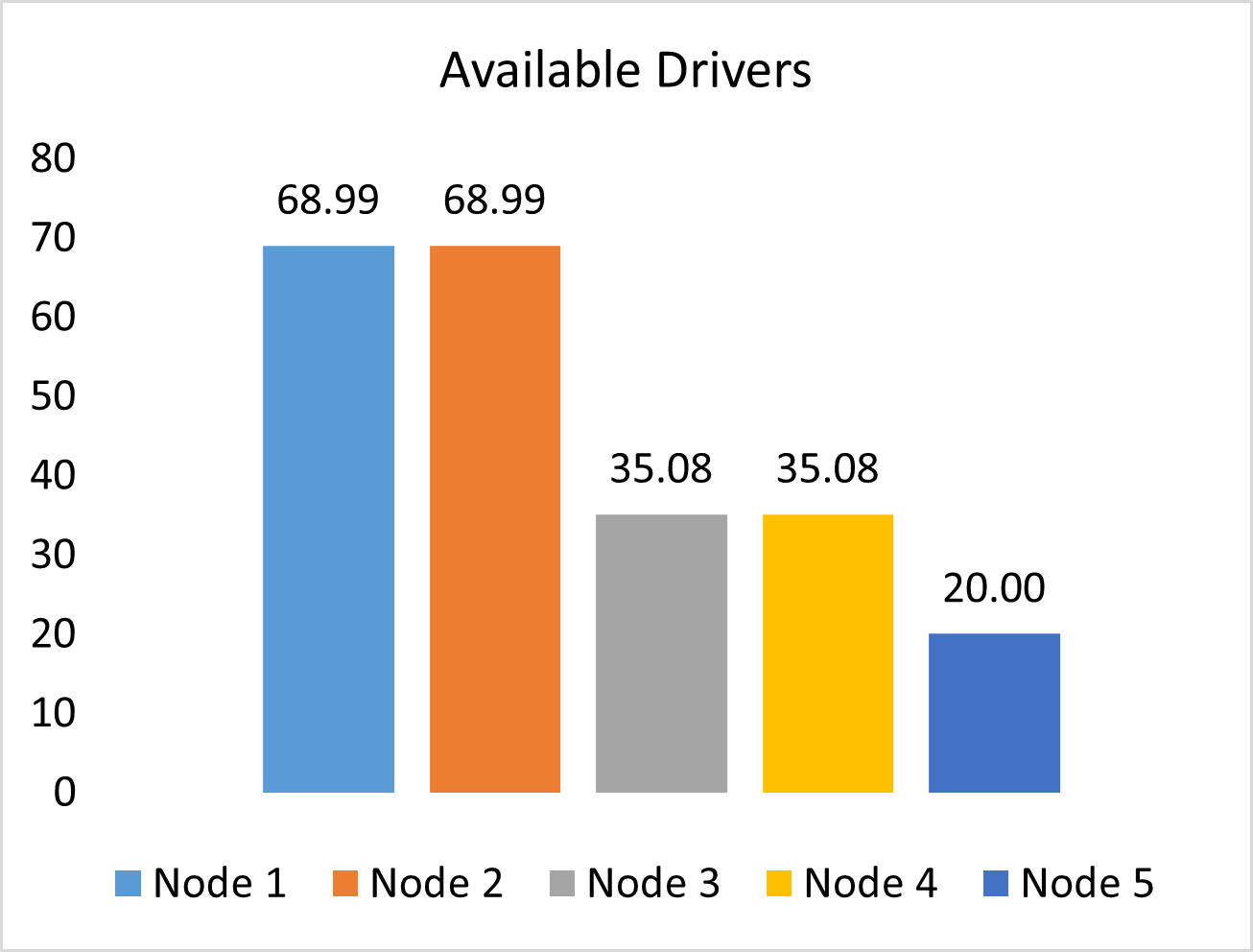}
        \caption{Available Drivers}
        \label{fig:Available Drivers}
    \end{subfigure}
    \hfill
    \begin{subfigure}[b]{0.49\textwidth}
        \centering
        \includegraphics[width=\textwidth]{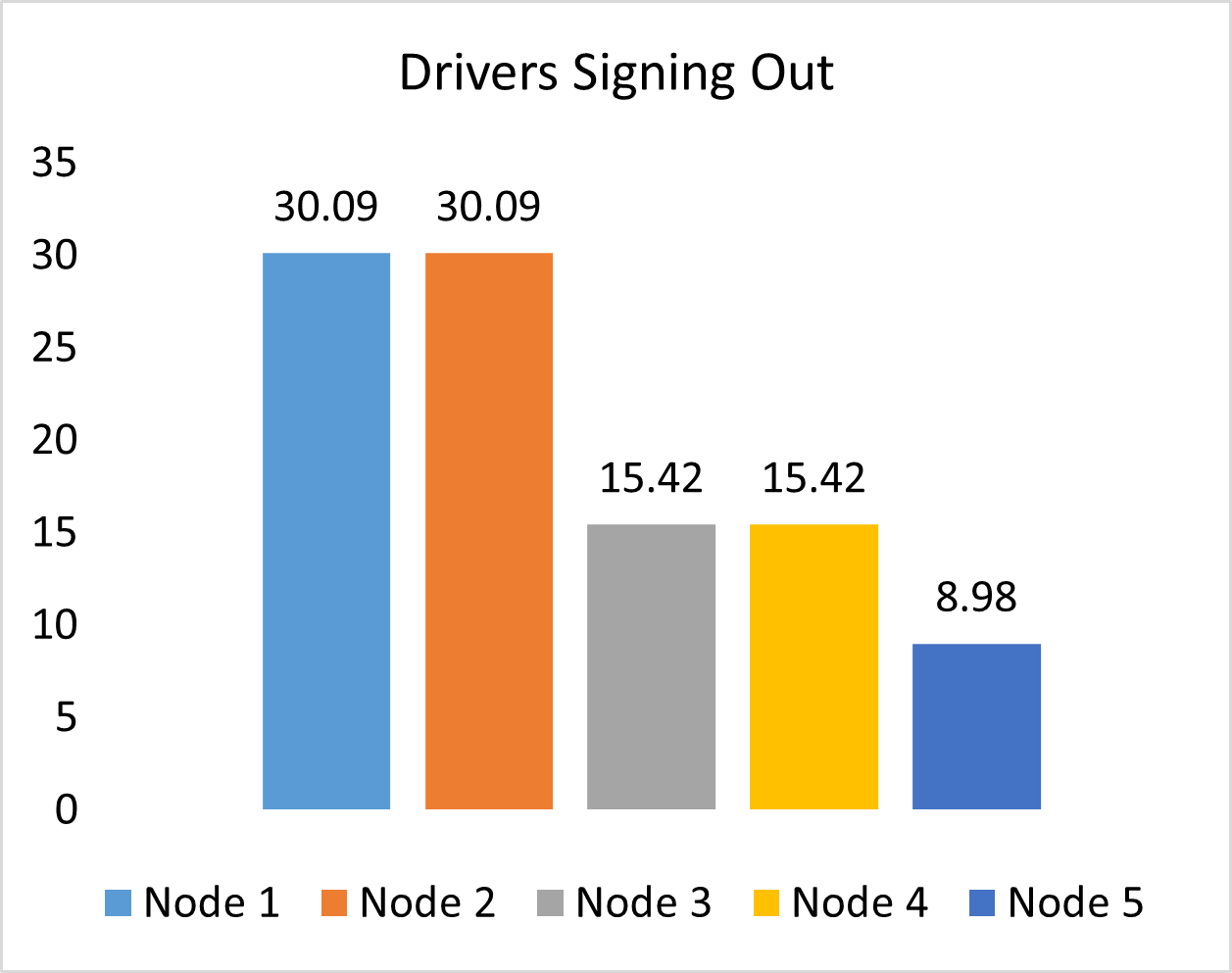}
        \caption{Driver's Signing Out}
        \label{fig:Driver's Signing Out}
    \end{subfigure}
    \caption{Available number of drivers and number of drivers signing out at equilibrium}
    \label{fig:flow1}
\end{figure}

Relocation time also explains the high number of available drivers at origin Node 1 and 2 shown in Figure \ref{fig:Available Drivers}. The number of drivers at the origin nodes is almost double compared to those at the transit hub. This is because there is no ride-sourcing demand from the transit hub to the destination as travelers only access transit there. Hence, drivers available at hubs either relocate to Node 1 and 2 or sign out. If ride-sourcing is used as a last-mile connection, drivers may stay at hubs and the number of available drivers would be higher. Node 5 has 20 available drivers at equilibrium, equal to the number signed in at Node 5, because Node 5 is neither a traveler's destination nor a transit hub.

From Figure \ref{fig:Driver's Signing Out}, we can see that locations having a high number of available drivers, i.e., Nodes 1 and 2,  have a high sign-out rate of drivers. This sign-out pattern of drivers among the nodes can be different if drivers consider an opportunity cost of ride-sourcing services. For instance, drivers may earn \$20 by signing out from the system and doing some other chores. In that case, the sign-out rate may be higher than the current value and their pattern may be different depending on the varying opportunity costs of ride-sourcing services among different nodes. We can include this opportunity cost in the driver's model \eqref{mod:driver} and reformulation model \eqref{mod:reformulation} without changing the overall modeling structure.

Another key output of this framework is the traveler's and driver's ride-sourcing prices, as shown in Figure \ref{fig:Traveler's and Driver's Price}, determined by the interaction of travelers and drivers' decision-making. Traveler's and driver's flow influence one another satisfying equilibrium constraints \eqref{eq:equilibrium}, and prices are dual variables of constraints \eqref{ride source flow to destination driver and rider balance} and \eqref{ride source flow to hub driver and rider balance} in model \eqref{mod:reformulation}. These prices are set relative to driving and transit costs. In equilibrium, multimodal prices are less than ride-sourcing prices, so that lower driver flow are attracted to transit hubs. Traveler's prices also follow this pattern because of low demand for multimodal services and shorter travel time to the hub.

Figure \ref{fig:Value of one extra driver} shows the penalty or value of one additional driver in the traveler's destination or transit hub. Our model estimates prices that balance supply-demand from the origin node and captures the marginal value of extra driver supply. According to Corollary~\ref{corollary 1}, ride-sourcing prices for OD (1,2) is the sum of the dual variables of constraint \eqref{ride source flow to destination driver and rider balance} for OD (1,2) and constraint \eqref{equilibrium constraint 3} for Node 2. Similarly, multimodal ride-sourcing prices for OD (1,2) is the sum of the dual variables of constraint \eqref{ride source flow to hub driver and rider balance} for OD (1,2) and constraint \eqref{equilibrium constraint 4} for Node 3, the transit hub for OD (1,2). In other words, traveler's prices include the costs for ride-sourcing services and the penalty/values of bringing an additional driver to their destinations or transit hubs. For example, if travelers request ride-sourcing to a remote area with limited returning trips, they may pay more to compensate for the impact of their travel. Depending on the sign of this value being positive (or negative), equilibrium traveler's price may be higher (or lower) than the corresponding driver's price.

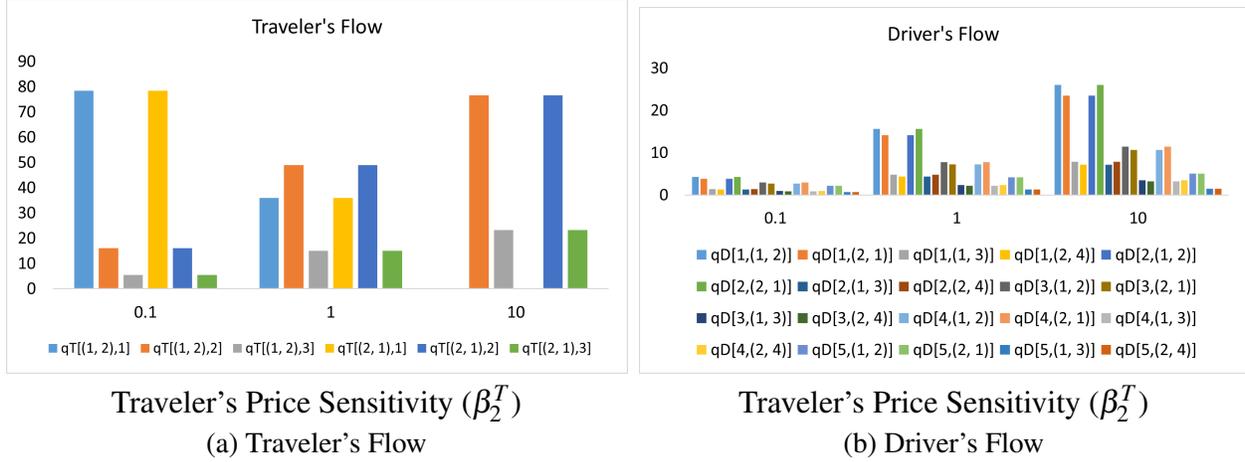
\begin{figure}[ht]
    \centering
    \begin{subfigure}[b]{0.50\textwidth}
        \centering
        \includegraphics[width=\textwidth]{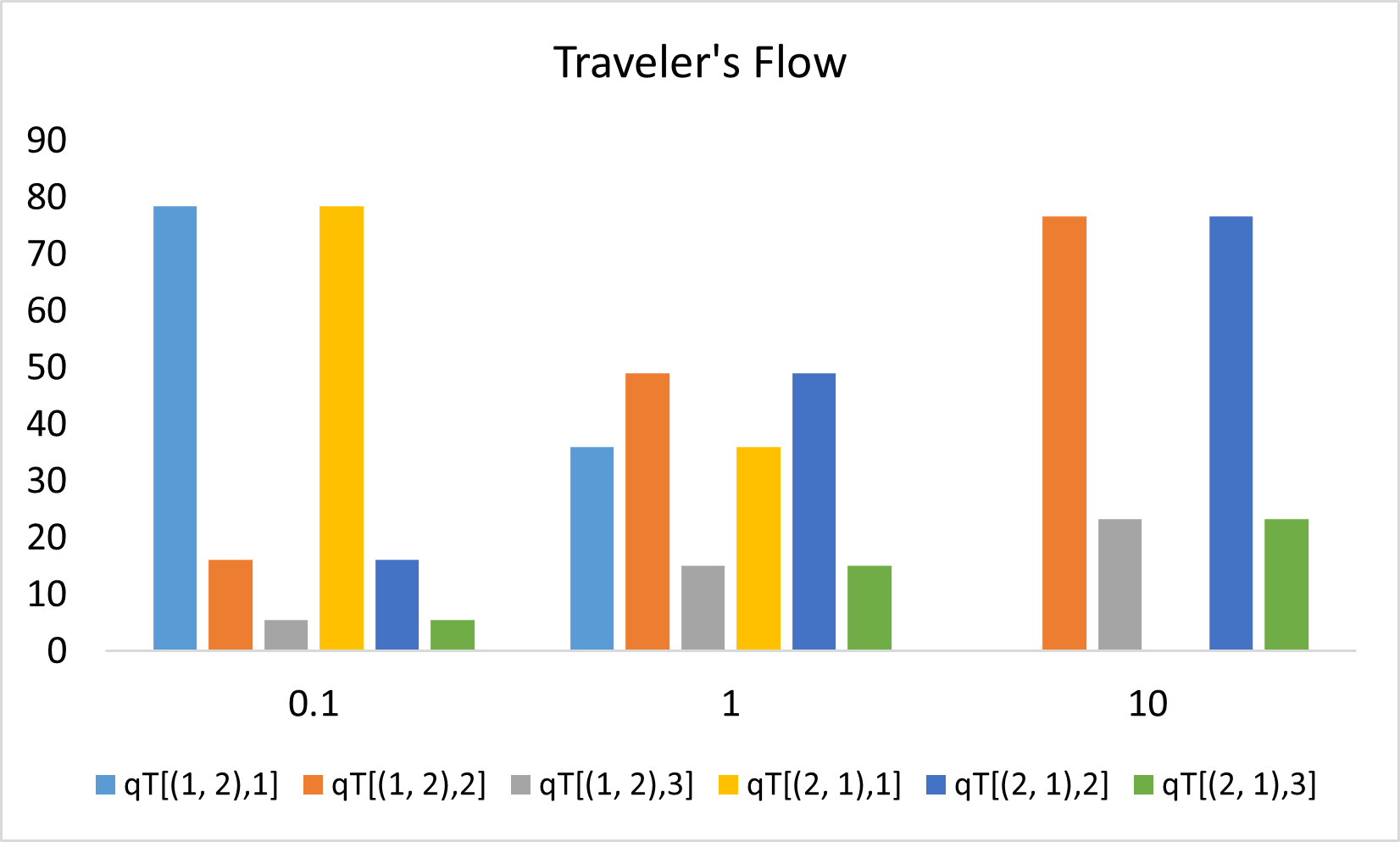}
        \label{fig:b2t_Traveler's Flow}
        \begin{tikzpicture}[overlay, remember picture]
            \node at (0, 0.1) {Traveler's Price Sensitivity ($\beta_2^T$)};
        \end{tikzpicture}
        \caption{Traveler's Flow}
    \end{subfigure}
    \hfill
    \begin{subfigure}[b]{0.49\textwidth}
        \centering
        \includegraphics[width=\textwidth]{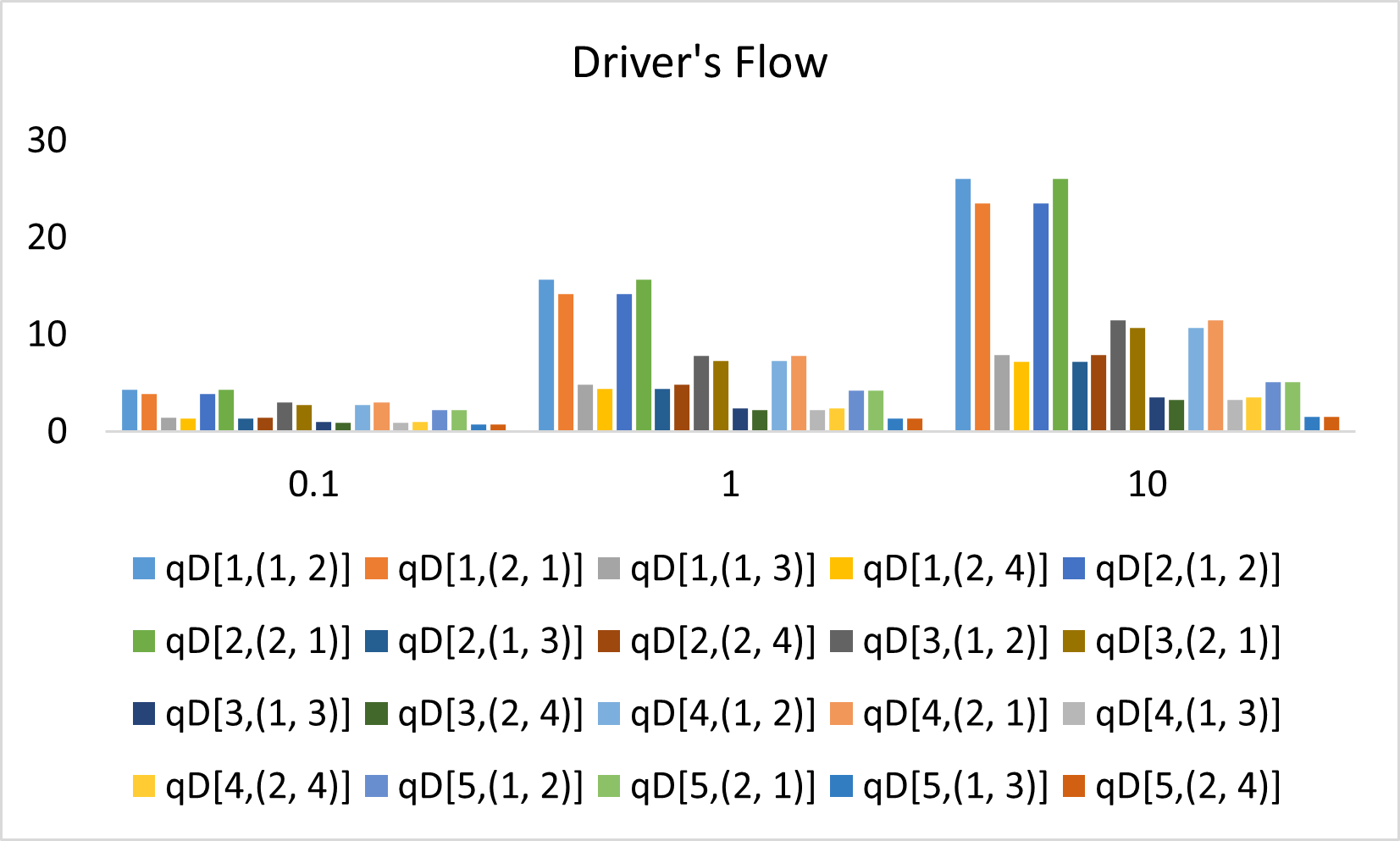}
        \label{fig:b2t_Driver's Flow}
        \begin{tikzpicture}[overlay, remember picture]
            \node at (0, 0.1) {Traveler's Price Sensitivity ($\beta_2^T$)};
        \end{tikzpicture}
        \caption{Driver's Flow}
    \end{subfigure}
    \caption{Changes of Traveler's and Driver's Flow with Traveler's Price Sensitivity}
    \label{fig:flow3}
\end{figure}

\begin{figure}[ht]
    \centering
    \begin{subfigure}[b]{0.50\textwidth}
        \centering
        \includegraphics[width=\textwidth]{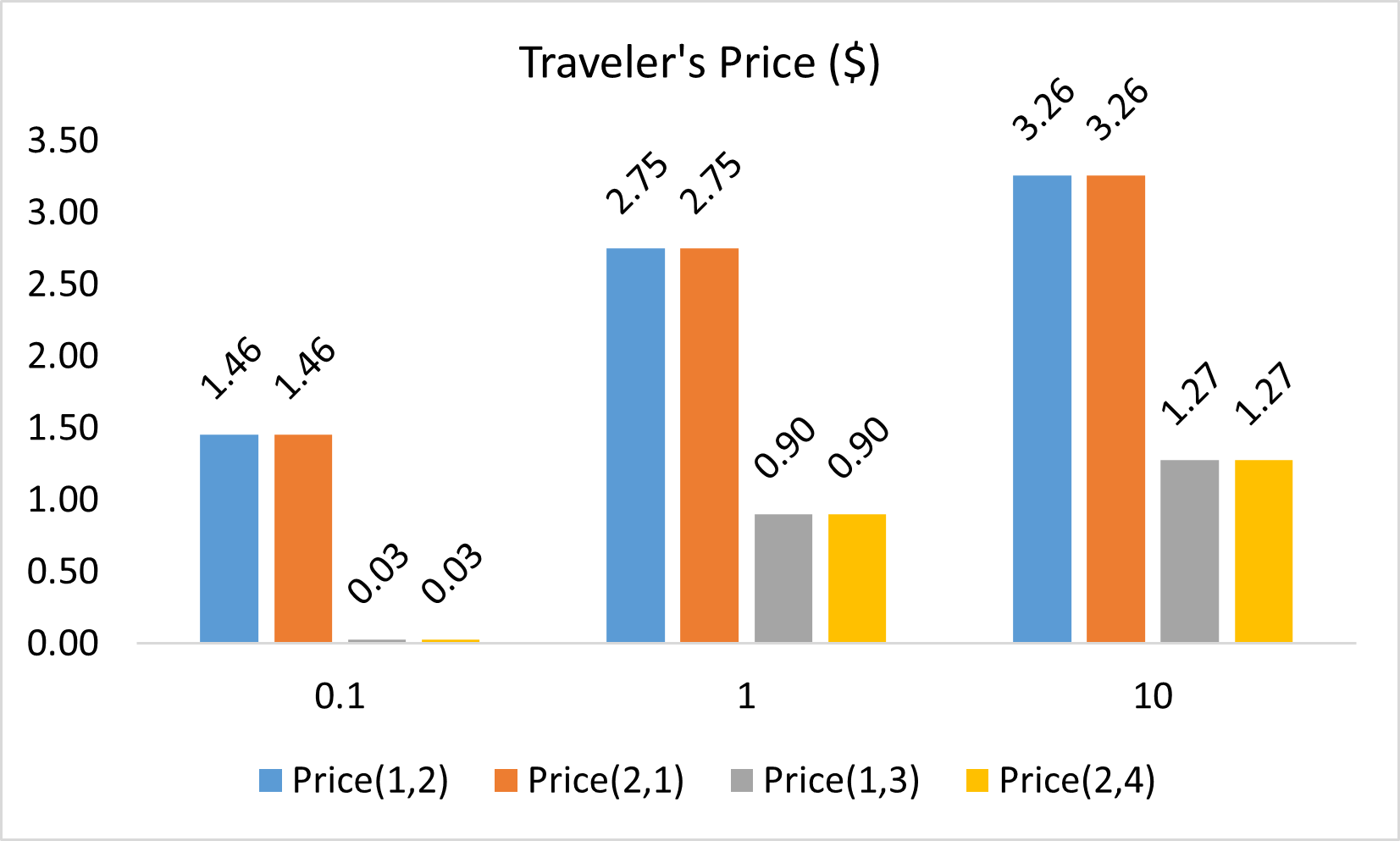}
        \label{fig:b2t_Traveler's Price}
        \begin{tikzpicture}[overlay, remember picture]
            \node at (0, 0.1) {Traveler's Price Sensitivity ($\beta_2^T$)};
        \end{tikzpicture}
        \caption{Traveler's Price}
    \end{subfigure}
    \hfill
    \begin{subfigure}[b]{0.49\textwidth}
        \centering
        \includegraphics[width=\textwidth]{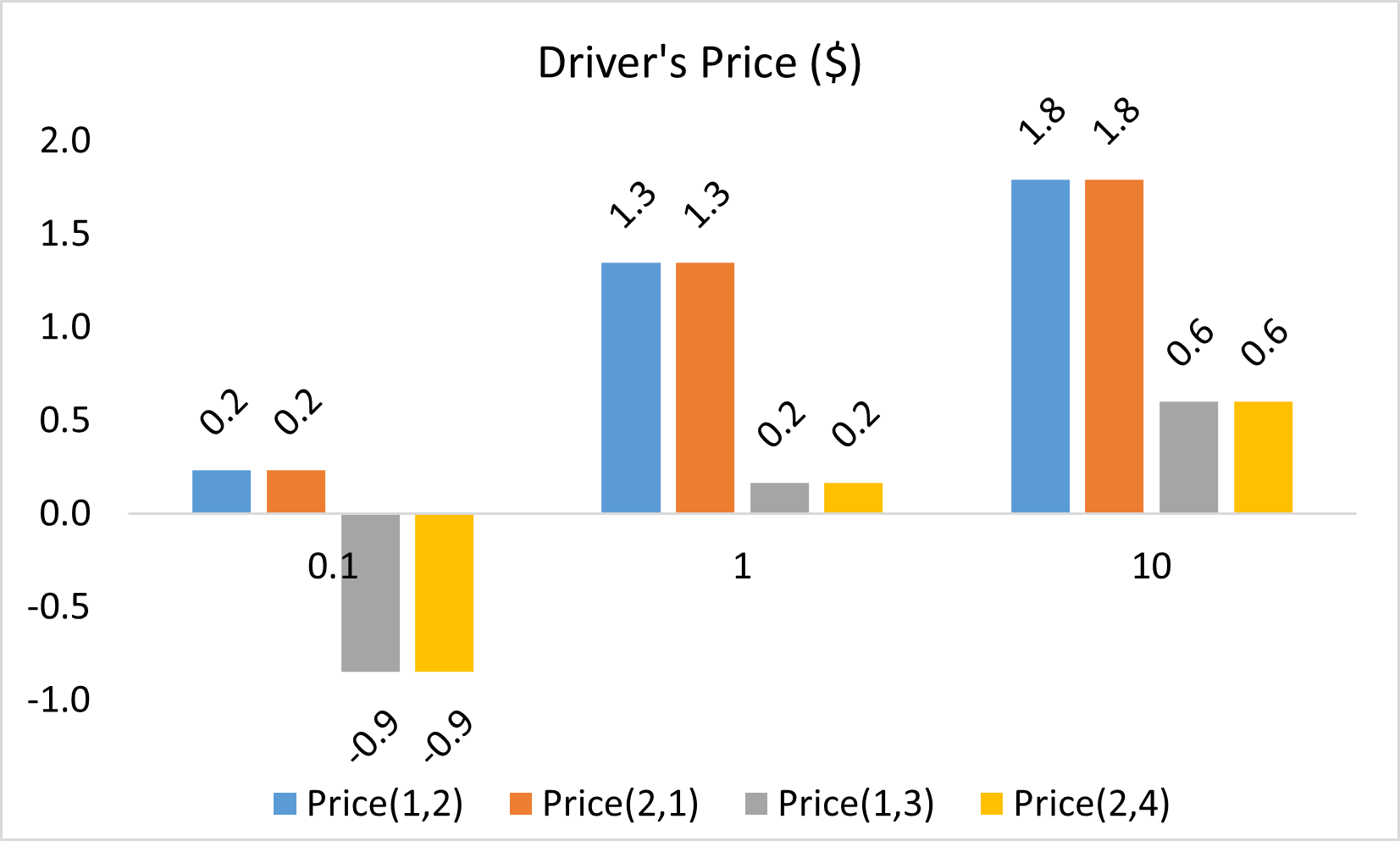}
        \label{fig:b2t_Driver's Price}
        \begin{tikzpicture}[overlay, remember picture]
            \node at (0, 0.1) {Traveler's Price Sensitivity ($\beta_2^T$)};
        \end{tikzpicture}
        \caption{Driver's Price}
    \end{subfigure}
    \hfill
    \begin{subfigure}[b]{0.49\textwidth}
        \centering
        \includegraphics[width=\textwidth]{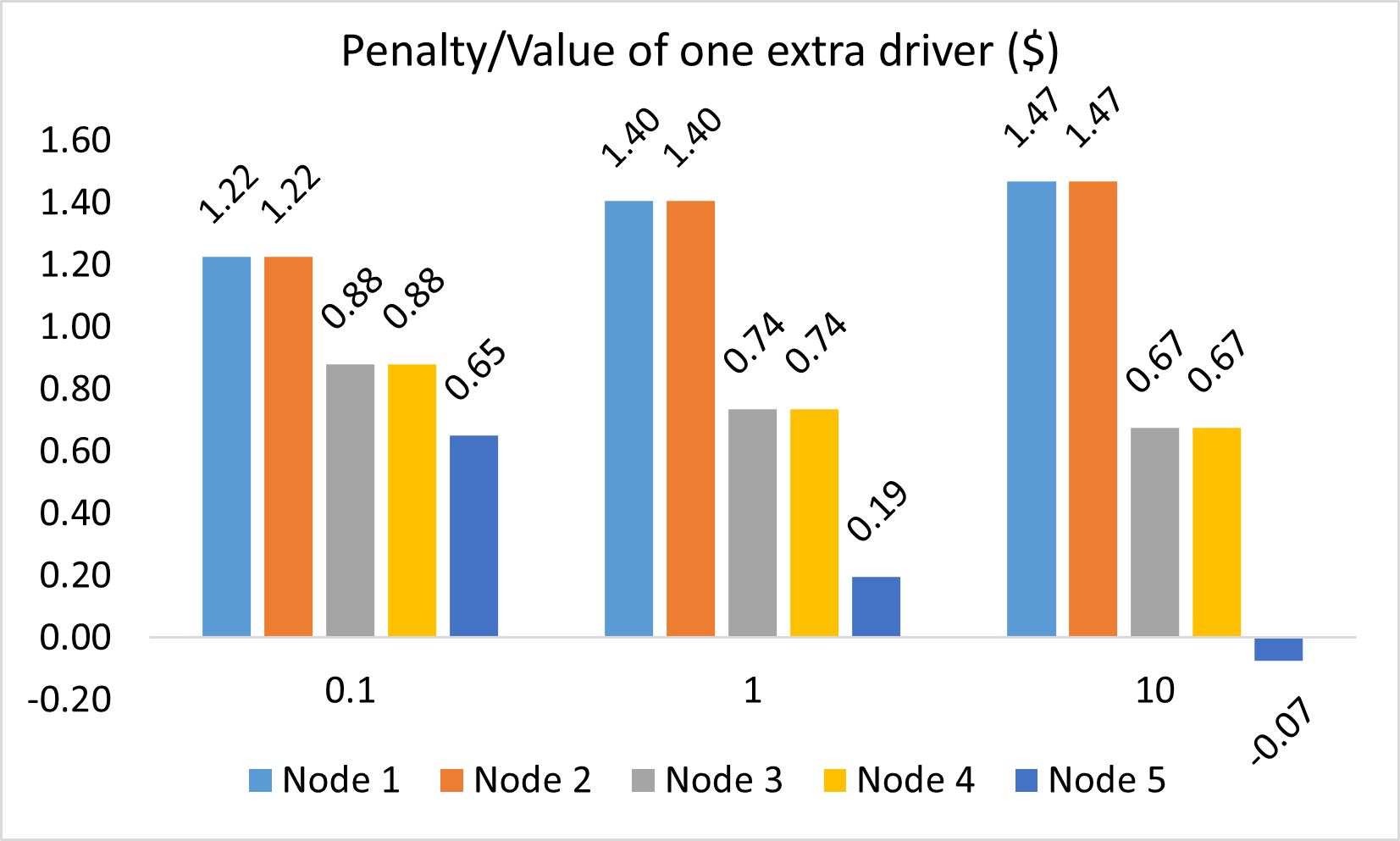}
        \label{fig:b2t_Value of one extra driver}
        \begin{tikzpicture}[overlay, remember picture]
            \node at (0, 0.1) {Traveler's Price Sensitivity ($\beta_2^T$)};
        \end{tikzpicture}
        \caption{Penalty/Value of adding an additional driver}
    \end{subfigure}
    \caption{Changes of Traveler's and Driver's Equilibrium Prices, and Penalty/Value of adding an additional driver with Traveler's Price Sensitivity}
    \label{fig:flow5}
\end{figure}
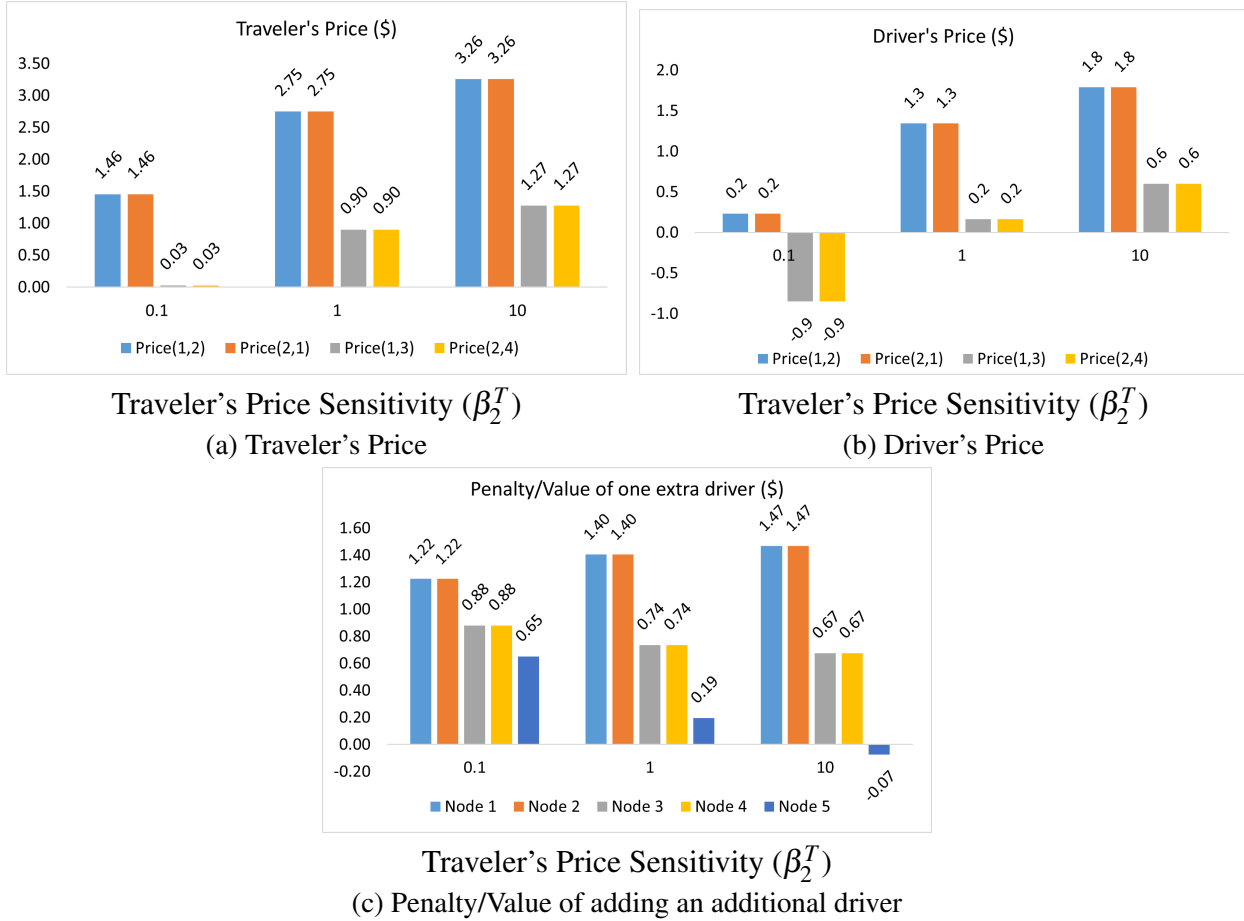

We change the price sensitivity of travelers to examine its effect on equilibrium prices. Figure \ref{fig:flow3}(a) shows that when sensitivity is low (0.1), most travelers choose high-priced driving mode. As sensitivity increases (1.0), most choose ride-sourcing, followed by driving and multimodal. Further increase of sensitivity (10) leads most travelers to use ride-sourcing, followed by multimodal, and almost no travelers use driving. This is because ride-sourcing and multimodal modes have lower prices (Figure \ref{fig:flow5}(a)) compared with driving, which includes both driving and parking costs. Even with sensitivity being high, ride-sourcing and multimodal costs are still lower than driving costs, making travelers more likely to choose those cheaper modes. As more travelers choose ride-sourcing and multimodal transportation at higher sensitivity, Figure \ref{fig:flow3}(b) shows that driver's flow also becomes higher.

Notice that drivers' prices serving the multimodal option are negative when travelers' price sensitivity is very low (Figure \ref{fig:flow5}(b)). This is because of the very low multimodal demand of travelers, resulting in a small flow of drivers serving multimodal transportation. As the payment of travelers is very low, to balance the low demand and make the system reach an equilibrium state, drivers serving multimodal transportation will have negative prices. Note that negative prices neither mean that there will be negative driver's payment in reality, nor it indicates any modeling flaw. Instead, it implies that a traveler's price sensitivity cannot be too small, which makes the whole system balanced in an unrealistic way. Again, because of this low number of drivers supply, the penalty or value of an additional driver at any location including the transit hub may increase which is evident from Figure \ref{fig:flow5}(c). As the driver's flow i.e. supply increases with higher price sensitivity, the penalty or value of adding one additional driver also decreases at all locations. When the driver's flow or the number of available drivers is too high at one location, the value of an additional driver there may also be negative, which implies that travelers may need to pay more to bring an additional driver to that location. Node 5 at high price sensitivity is an example in this case. As Node 5 is neither a traveler destination nor a transit hub, drivers have the only option to relocate to a zone where there is travel demand. There is no traveler flow to balance the high number of drivers directly, leading to a negative value of any additional driver at that location. 

\subsection{Sioux Falls Network}
The Sioux Falls test network has 24 nodes and 76 links. With the purpose of drawing policy insights, we have considered 7 origins ($\mathcal{R}=[1,4,5,6,7,19,23]$) and 7 destinations ($\mathcal{S}=[13,24,22,21,20,5,9]$) for travelers. Hence, travelers OD pairs in the network, \[
\mathcal{RS}=\{(1, 13), (4, 24), (5, 22), (6, 21), (7, 20), (19, 5), (23, 9)\}
\] Ride-sourcing drivers may be available at any nodes of the network and may (or may not) relocate to origin node to serve the travelers.

\begin{figure}[ht]
    \centering
    \begin{subfigure}[b]{0.32\textwidth}
        \centering
        \includegraphics[width=\textwidth]{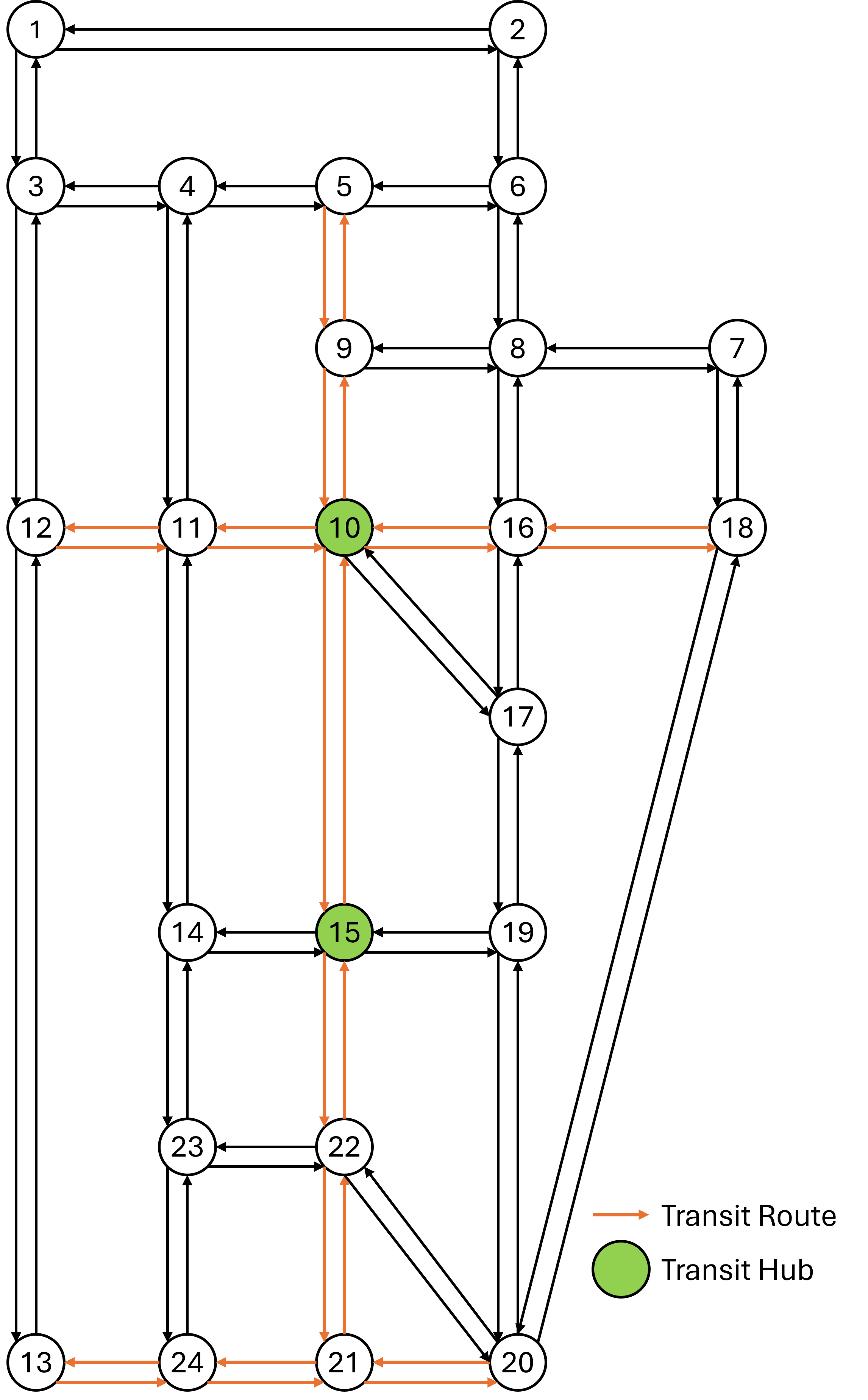}
        \label{fig:sx_2 hub}
        \caption{scenario (1): 2 Transit Hub}
    \end{subfigure}
    \hfill
    \begin{subfigure}[b]{0.32\textwidth}
        \centering
        \includegraphics[width=\textwidth]{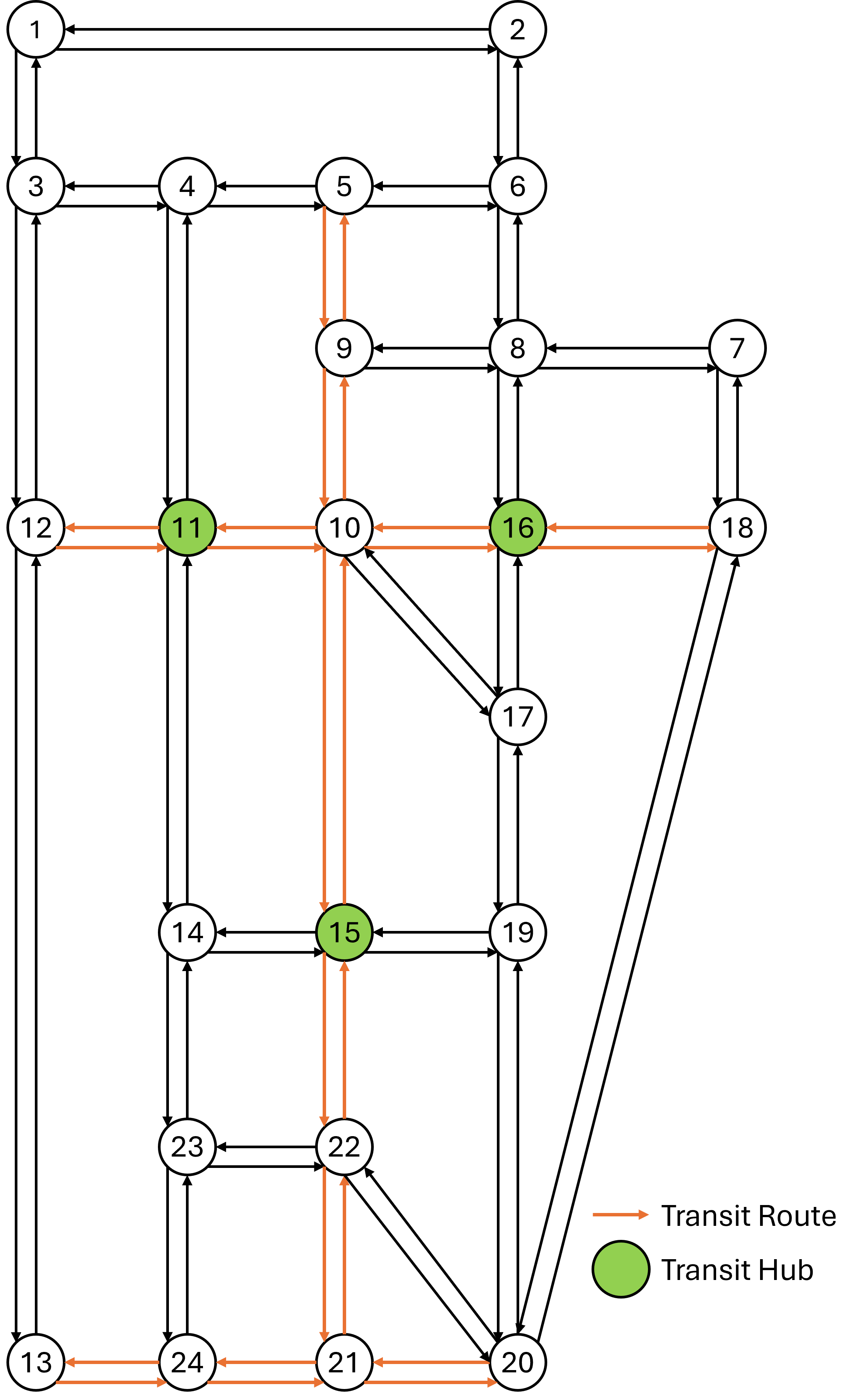}
        \label{fig:sx_3 hub}
        \caption{scenario (2): 3 Transit Hub}
    \end{subfigure}
    \hfill
    \begin{subfigure}[b]{0.32\textwidth}
        \centering
        \includegraphics[width=\textwidth]{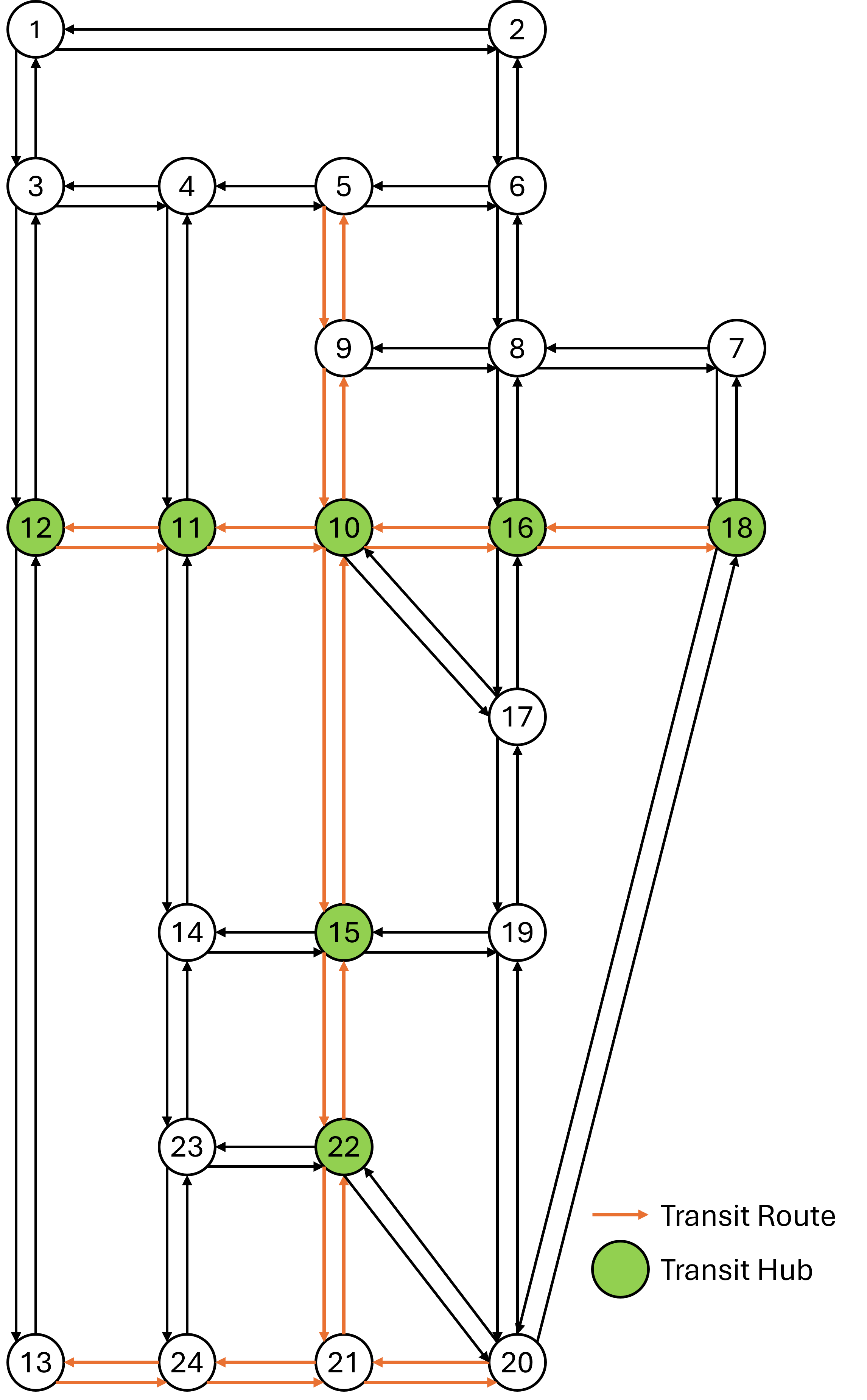}
        \label{fig:sx_7 hub}
        \caption{scenario (3): 7 Transit Hub}
    \end{subfigure}
    \caption{Sioux Falls Test Network}
    \label{fig:flow15}
\end{figure}

We conduct a case study to understand how the number of transit hub in the network may effect the MT system. We consider three scenarios, where we change the number of transit hubs in the network for each scenario to test it's impact on the MT system. The scenarios are: (1) 2 transit hubs, $\mathcal{H}=[10,15]$; (2) 3 transit hubs, $\mathcal{H}=[11,16,15]$; and (3) 7 transit hubs, $\mathcal{H}=[12,11,10,16,18,15,22]$ (Table~\ref{tab:Mapping of OD with Transit Hub}). Notice that for scenario (1) and (2), the transit hubs are shared among multiple OD pairs. The mapping of each transit hub corresponding to the travelers OD are given in Table~\ref{tab:Mapping of OD with Transit Hub}. Driver's OD set $\mathcal{\overline{RS}}$ not only includes travelers ride-sourcing OD $\mathcal{RS}$, but also consists of OD connecting travelers origin to corresponding transit hub. The transit routes are considered fixed for all scenarios and shown in Figure \ref{fig:flow15}. The location of transit hubs for each scenarios are also shown in Figure \ref{fig:flow15}. The sensitivity parameters are considered same as the '5 Node Test Network'. The travel time and relocation time considered for each OD are deduced using the free-flow travel time (See \cite{stabler2021}) for each link along the shortest path. The travel demand of the OD pairs are for all scenarios: 
$\mathcal{RS}[1,13] = 500$, $\mathcal{RS}[4,24] = 200$, $\mathcal{RS}[5,22] = 200$, $\mathcal{RS}[6,21] = 100$, $\mathcal{RS}[7,20] = 500$, $\mathcal{RS}[19,5] = 100$, $\mathcal{RS}[23,9] = 500$ \citep{stabler2021}. Travel time through transit is considered double of the free-flow travel time along the transit route.

\begin{table}[ht]
\centering
\caption{Mapping of OD with Transit Hub}
\begin{tabularx}{\textwidth}{|X|X|X|}
\hline
\textbf{Scenario 1} & \textbf{Scenario 2} & \textbf{Scenario 3} \\
\hline
No. of Transit Hub = 2 & No. of Transit Hub = 3 & No. of Transit Hub = 7 \\
\(\mathcal{H} = [10,15]\) & \(\mathcal{H} = [11,16,15]\) & \(\mathcal{H} = [12,11,10,16,18,15,22]\) \\
 & & \\
Mapping of OD with Hub: & Mapping of OD with Hub: & Mapping of OD with Hub: \\
\(h(1,13) = 10\) & \(h(1,13) = 11\) & \(h(1,13) = 12\) \\
\(h(4,24) = 10\) & \(h(4,24) = 11\) & \(h(4,24) = 11\) \\
\(h(5,22) = 10\) & \(h(5,22) = 11\) & \(h(5,22) = 10\) \\
\(h(6,21) = 10\) & \(h(6,21) = 16\) & \(h(6,21) = 16\) \\
\(h(7,20) = 10\) & \(h(7,20) = 16\) & \(h(7,20) = 18\) \\
\(h(19,5) = 15\) & \(h(19,5) = 15\) & \(h(19,5) = 15\) \\
\(h(23,9) = 15\) & \(h(23,9) = 15\) & \(h(23,9) = 22\) \\
\hline
\end{tabularx}
\label{tab:Mapping of OD with Transit Hub}
\end{table}

\begin{figure}[ht]
    \centering
    \begin{subfigure}[b]{0.32\textwidth}
        \centering
        \includegraphics[width=\textwidth]{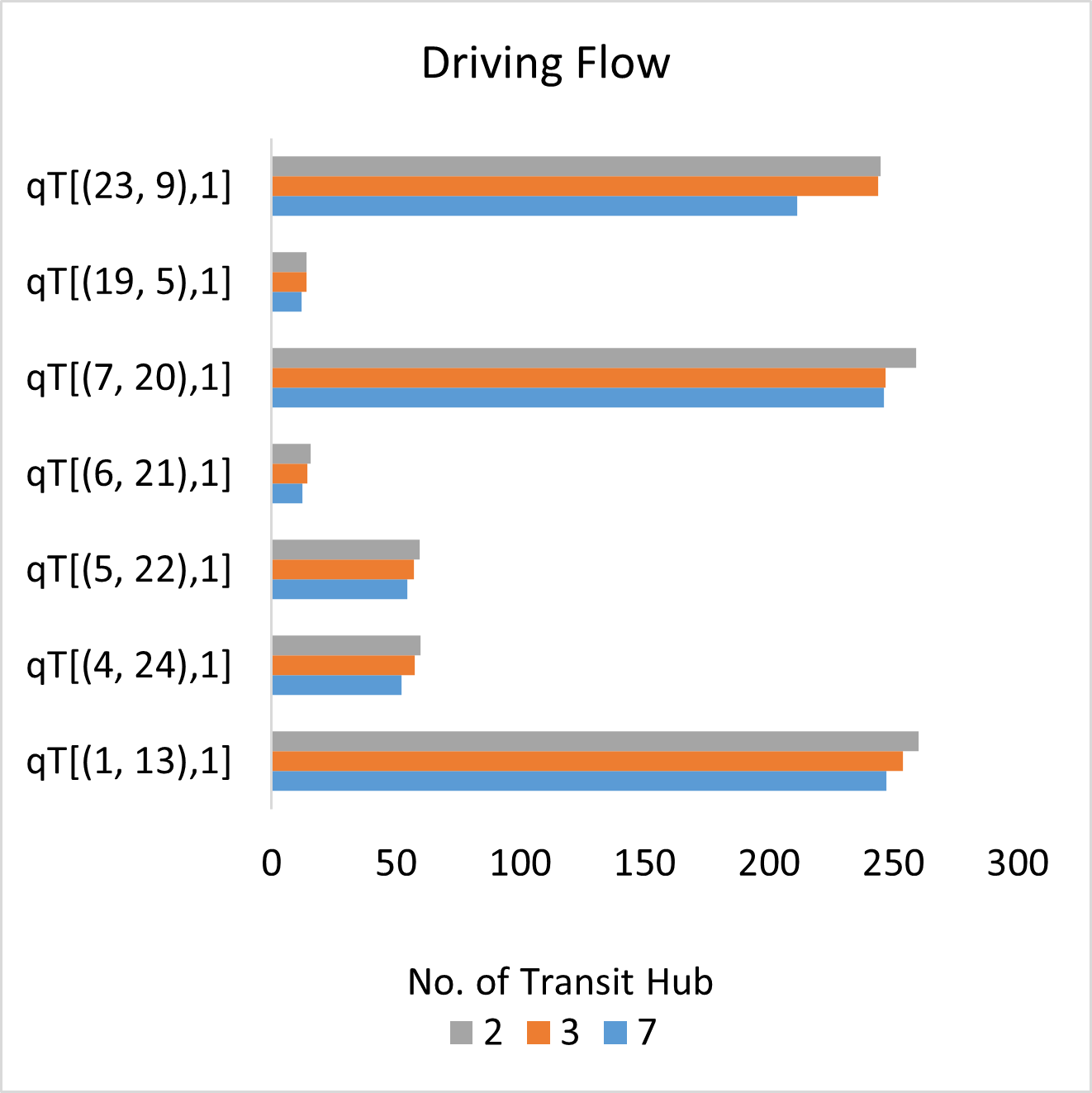}
        \label{fig:sx_tr_dr_flow}
        \caption{Driving Flow}
    \end{subfigure}
    \hfill
    \begin{subfigure}[b]{0.32\textwidth}
        \centering
        \includegraphics[width=\textwidth]{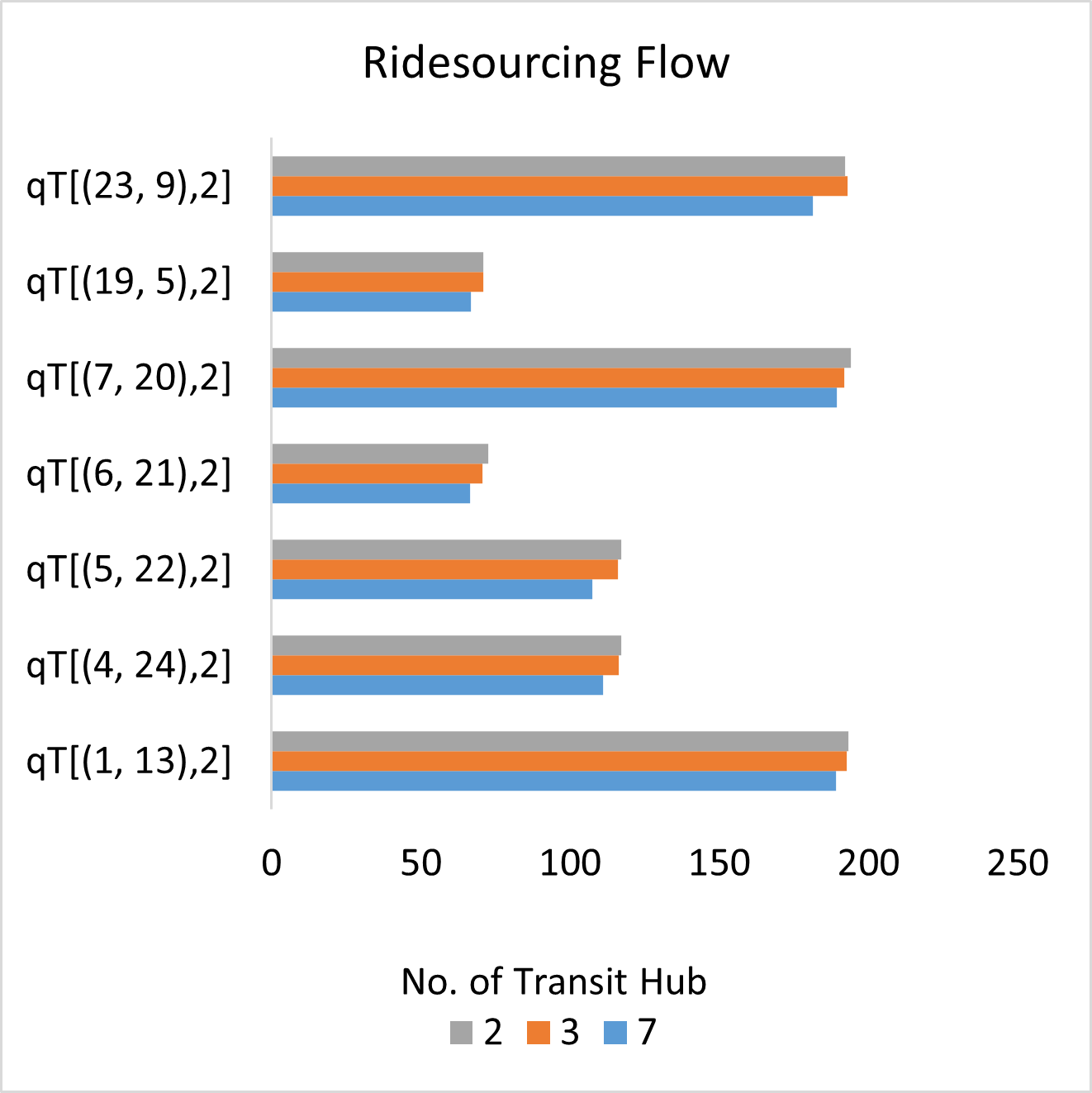}
        \label{fig:sx_tr_ride_flow}
        \caption{Ride-sourcing Flow}
    \end{subfigure}
    \hfill
    \begin{subfigure}[b]{0.32\textwidth}
        \centering
        \includegraphics[width=\textwidth]{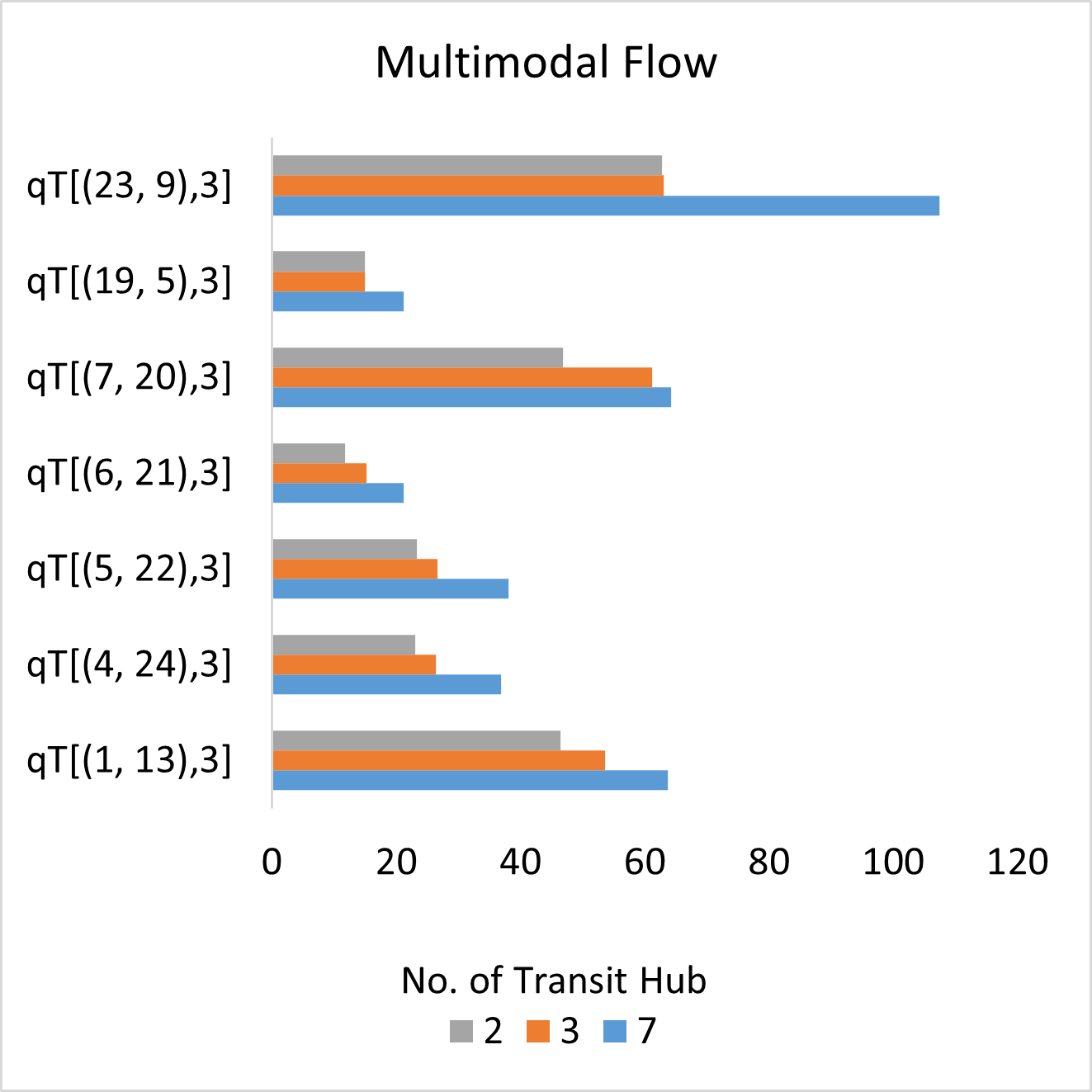}
        \label{fig:sx_tr_multi_flow}
        \caption{Multimodal Flow}
    \end{subfigure}
    \caption{Change of Traveler's flow with number of transit hub}
    \label{fig:flow16}
\end{figure}

\begin{figure}[ht]
    \centering
    \begin{subfigure}[b]{0.49\textwidth}
        \centering
        \includegraphics[width=\textwidth]{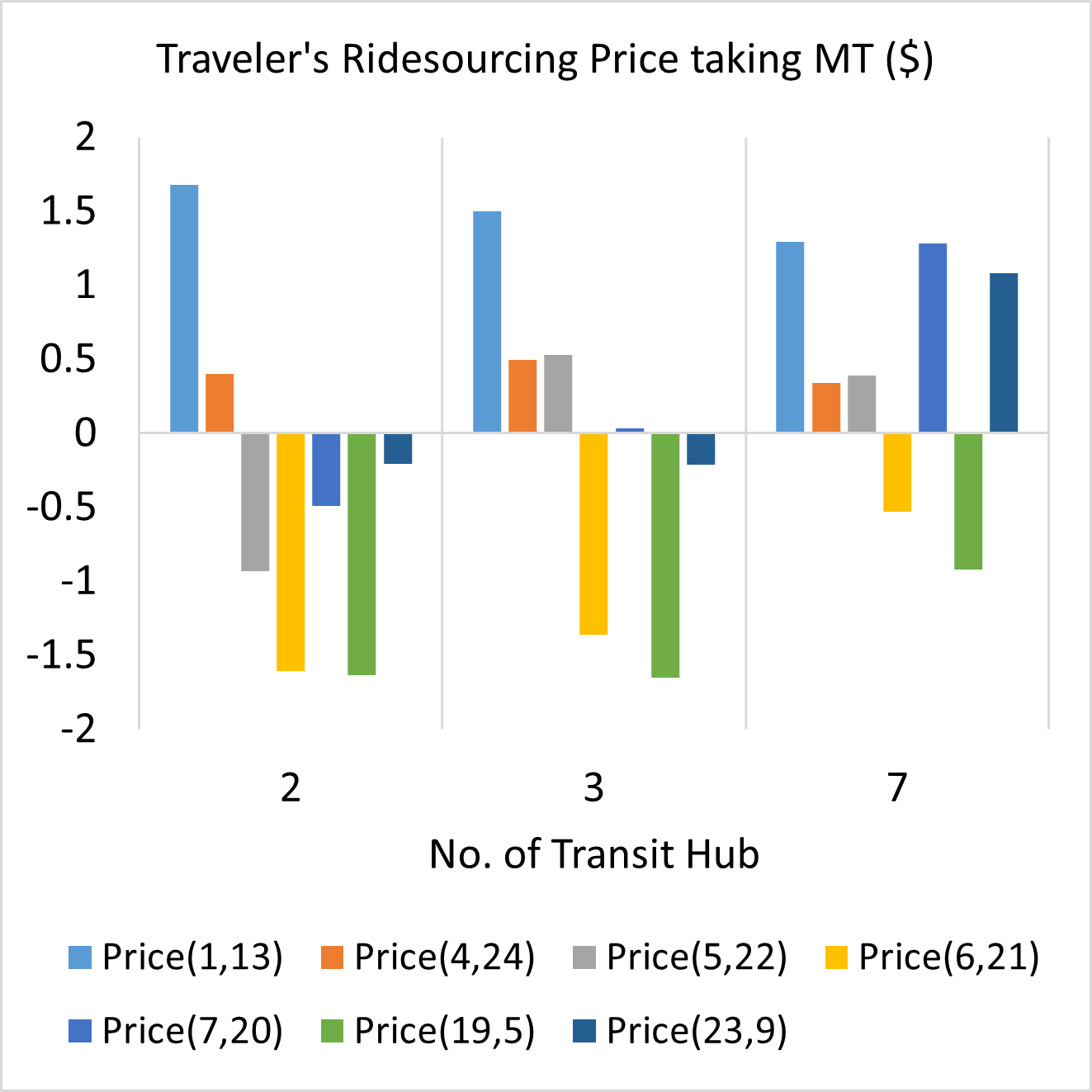}
        \label{fig:sx_tr_mt_price}
        \caption{Traveler's ride-sourcing price serving MT}
    \end{subfigure}
    \hfill
    \begin{subfigure}[b]{0.49\textwidth}
        \centering
        \includegraphics[width=\textwidth]{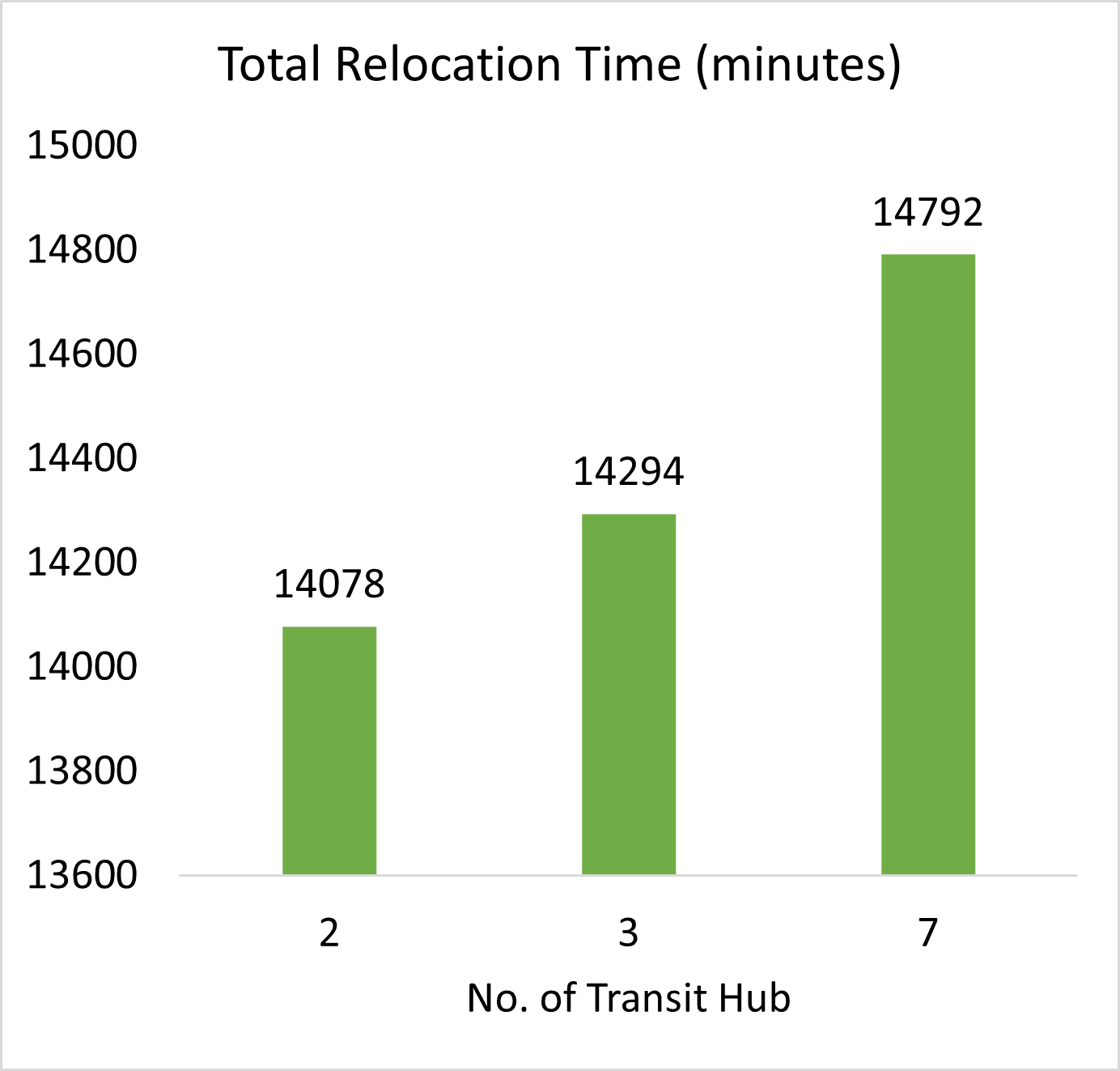}
        \label{fig:tot_rel_time}
        \caption{Total relocation time of drivers}
    \end{subfigure}
    \caption{Traveler's ride-sourcing price serving MT and total relocation time of drivers}
    \label{fig:flow17}
\end{figure}

Figure \ref{fig:flow16}(a) shows that traveler's flow through driving decreases when the number of transit hub in the network is more. Traveler's ride-sourcing flow also decreases with the increase of the transit hub for most of the OD (Figure \ref{fig:flow16}(b)). But Figure \ref{fig:flow16}(c) show that the multimodal flow increases with the increase of transit hub. It indicates that travelers prefer the MT more than other existing travel modes if they are facilitated by more transit hubs in the network. More transit hub will increase the accessibility of transit service from any location, hence traveler can easily reach to the transit station using the first-mile ride-sourcing service at a less amount of time. 

As evident from Figure \ref{fig:flow17}(a), there is no consistent pattern for traveler's price of ride-sourcing serving the MT. Though travel time of the first-mile ride-sourcing decreases for increased number of transit hub, different configuration of transit hub location may impose additional impact on the traveler's price. Despite having any noticeable pattern, we find that traveler's price may become negative for most of the OD if there is fewer transit hub i.e., they may need to be subsidized by the TNCs or society to use the MT. As TNCs regulate the ride-sourcing drivers in practice, it may reduce the profit of TNCs if they wish to serve the MT in a transportation network with fewer transit hubs. It iterates the need of coordination between TNCs and transit agency though both of them have their own selfish objectives. Transit agency may need to plan the location of transit hubs by coordinating with TNCs if they wish to mitigate their first-mile connection problem and increase transit riderships.

Although more number of transit hubs may increase the MT ridership in the network, it may influence the ride-sourcing driver's total relocation time serving direct ride-sourcing and MT services. Figure \ref{fig:flow17}(b) show that their total relocation time in the network increase with increasing number of transit hubs. As congestion is not considered in the network and relocation time of drivers from any node to traveler's origin is fixed, higher MT demand may be the only cause of driver's higher relocation time. When demand for MT become large, more flow of ride-sourcing drivers are needed to balance the high demand, which eventually increases their total relocation time in the network. Note that relocation time is also related with empty vehicle miles traveled (VMT) for ride-sourcing drivers \cite{JIA2022103318, fakhrmoosavi2024}. The increase of relocation time would also increase the empty VMT of drivers, which actually reduces the efficiency of the proposed MT system. Higher empty VMT would also contribute for higher emission from transportation sector and it would defeat the sustainability goal of the MT system \cite{JIA2022103318}. If ride-sourcing demand exists from transit hubs to any locations, drivers may not need to relocate to provide service. It would eventually reduce the empty VMT and make the MT system more sustainable.

\section{Conclusion} \label{sec:conclusion}
In this research, we have developed a unified modeling framework leveraging multi-agent optimization to capture the non-cooperative behavior among travelers and drivers in a multimodal system. We developed effective computational strategies for solving the multi-agent optimization problem of multimodal systems. We show that the problem is equivalent to solving a single reformulated convex optimization problem, which can be efficiently solved to global optimum using commercial non-linear solvers. We also proved the existence and uniqueness of solutions for such a problem.

We determine the traveler's and driver's flow that are balanced at each zone over the transportation network as well as the number of available drivers at a certain location in an equilibrium state. Our model is able to output locational equilibrium prices that balance each transportation mode. The equilibrium prices not only reflect the prices of the ride-sourcing services for unimodal and multimodal transportation but also capture the value/penalty of an additional driver available at each node. Such information is helpful to guide the quantitative incentive design to better promote sustainable transportation. We find that (1) if travelers are more sensitive to prices, they prefer ride-sourcing and multimodal transportation more than the driving option. As a result, more ride-sourcing drivers would be required and equilibrium ride-sourcing prices would increase to balance the increased demand. (2) Travelers may need to be subsidized by TNC or society to use multimodal services if ride-sourcing drivers become too sensitive to the prices. (3) For a location that is not a destination of any traveler or a transit hub, the value/penalty of an additional driver at that location is smaller than that value at a traveler's destination or transit hub. (4) Travelers may need to be subsidized to use multimodal transportation if the number of transit hub in the network is less. (5) While more transit hubs in the network may increase the demand for multimodal transportation, it would also increase the total empty VMT of ride-sourcing drivers by increasing the total relocation time.

Our research can be extended in several directions. First of all, we ignore the effect of ride-sourcing waiting time on the decision-making process of travelers and ride-sourcing drivers. This type of model is appropriate when waiting time is an insignificant factor in the decision-making process of agents. For instance, for long-term transportation planning problems in a large network, travelers' travel time and drivers' relocation time can significantly exceed the waiting time, which is negligible in the decision-making process. In addition, travelers intending to use ride-sourcing services may request a ride earlier than their intended trip start time, which would eventually offset the effect of waiting time in the traveler's mode choice decision-making process \cite{AFIFAH2022103777}. However, the proposed modeling framework can be extended to incorporate the impacts of waiting time by adding corresponding utility terms of waiting time in the traveler's and driver's decision-making process. Further, congestion is not considered in our model. We can extend our model for a congested transportation network by considering the traffic user equilibrium along with the discrete choice model. Additionally, our study only captures the interaction between travelers and drivers providing ride-sourcing services in a multimodal transportation system. One can further consider TNCs and public transit agencies as additional agents in the modeling framework.

\textbf{Acknowledgement}\\
This research is supported by National Science Foundation (NSF) under Grant No. 2041446.

\newpage

\bibliographystyle{trb}
\bibliography{trb_template}
\end{document}